\documentclass[a4paper,12pt]{article}
\usepackage{multirow}
\usepackage{graphicx,graphics}
\usepackage{amssymb}
\usepackage{array}
\usepackage{color}
\usepackage{graphics,graphicx,amssymb}

\begin{document}
 
\begin{flushright}
ADP-16-17/T972
\end{flushright}

\thispagestyle{empty}

\begin{center}

\begin{center}

\vspace{1.7cm}

{\LARGE\bf Optimising Charged Higgs Boson Searches at the Large Hadron Collider Across $b\bar b W^\pm$ Final States}
\end{center}

\vspace{1.4cm}

\renewcommand{\thefootnote}{\fnsymbol{footnote}}
{\bf Stefano Moretti$^{\,1\,}$}\footnote{E-mail: \texttt{s.moretti@soton.ac.uk}},
{\bf Rui Santos$^{\,2,\,3\,}$}\footnote{E-mail: \texttt{rasantos@fc.ul.pt}} and
{\bf Pankaj Sharma$^{\,4\,}$}\footnote{E-mail: \texttt{pankaj.sharma@adelaide.edu.au}}
\\

\vspace{1.cm}

${}^1\!\!$
{\em School of Physics and Astronomy, University of Southampton,} \\
{\em Southampton, SO17 1BJ, United Kingdom}
\\\vskip 10pt
${}^2\!\!$
{\em Centro de F\'{\i}sica Te\'{o}rica e Computacional,
    Faculdade de Ci\^{e}ncias,
    Universidade de Lisboa,} \\
{\em Campo Grande, Edif\'{\i}cio C8 1749-016 Lisboa, Portugal}
\\\vskip 10pt
${}^3\!\!$
{\em {ISEL -
 Instituto Superior de Engenharia de Lisboa,\\
 Instituto Polit\'ecnico de Lisboa
 1959-007 Lisboa, Portugal}
}\\\vskip 10pt

${}^3\!\!$
{\em {CoEPP -
 Center of Excellence in Particle Physics at Tera Scale,\\
 The University of Adelaide,
 5005 Adelaide, South Australia}
}\\

\end{center}

\begin{abstract}
In the light of the most recent data from Higgs boson searches and analyses, we re-assess the scope of the Large Hadron Collider in accessing heavy charged Higgs boson signals in $b\bar b W^\pm$ final states, wherein the contributing channels can be $H^+\to t\bar b$, $hW^\pm, HW^\pm$ and $AW^\pm$. We consider a 2-Higgs Doublet Model Type-II and we assume as  production mode $bg\to tH^-$ + c.c., the dominant one over the range $M_{H^\pm}\ge 480$ GeV, as dictated by $b\to s\gamma$ constraints. Prospects of detection are found to be significant for various Run 2 energy and luminosity options. 
\end{abstract}

\newpage
\setcounter{page}{1}
\setcounter{footnote}{0}

\section{Introduction} 
The discovery of a (singly) charged Higgs boson would signal the existence of a second Higgs doublet in addition to the Standard Model (SM)-like one already established through the discovery of the $W^\pm$ and $Z$ bosons at the S$p\bar p$S in the eighties and of a Higgs boson itself at the LHC only four years ago. Such a scalar field can naturally be accommodated in 2-Higgs Doublet Models (2HDMs). In its CP-conserving versions, they present in their spectra, after spontaneous Electro-Weak Symmetry Breaking (EWSB), five physical Higgs states: the neutral pseudoscalar ($A$), the lightest ($h$) and heaviest ($H$) neutral scalars and two charged ones ($H^\pm$). 

Of all 2HDM Yukawa types (see Ref.~\cite{Branco} for a review), we concentrate here on the 2HDM Type II (2HDM-II). Herein, constraints from $b\to s\gamma$ decays put a lower limit on the $H^\pm$ mass at about 480 GeV, rather independently of $\tan\beta$~\cite{Misiak:2015xwa}, the ratio of the Vacuum Expectation Values (VEVs) of the two doublets. Such a heavy mass region is very difficult to access because of the large reducible and irreducible backgrounds associated with the main decay mode $H^+\to t\bar b$, following the dominant production channel $bg\to t H^-$ \cite{bg}. (Notice that the rate of the latter exceeds by far other possible production modes \cite{bq}--\cite{ioekosuke}, thus rendering it the only viable channel at the CERN machine in the heavy mass region.) The analysis of the $H^+\to t\bar b$ signature has been the subject of many early debates \cite{roger}--\cite{roy1}, their conclusion being that  the LHC discovery potential might be satisfactory, so long that $\tan\beta$ is small ($\leq 1.5$) or large ($\geq30$) enough and the charged Higgs boson mass is below 600 GeV or so. Such positive prospects have very recently been revived by an ATLAS analysis of the full Run 1 sample \cite{ATLAS}, which searched precisely for the aforementioned $H^\pm$ production and decay modes, by exploring the mass range 200 to 600 GeV using multi-jet final states with one electron or muon. This study used multivariate analysis techniques in the signal-rich region while it employed control regions to reduce the large uncertainties on the backgrounds. An excess of data with respect to the SM predictions was observed for all $H^\pm$ mass hypotheses up to (but excluding) 600 GeV.  While CMS does not confirm such an excess \cite{CMS}, the increased sensitivity that the two experiments are accruing with the first Run 2 data calls for a renewed interest
in this respect.

In this spirit, and recognising that the $H^+\to t\bar b$ decay channel eventually produces a $b\bar b W^+$ signature, Ref.~\cite{Uppsala} attempted to extend the reach afforded by this channel by exploiting the companion signature $H^+\to h_{\rm SM}W^+$ $\to$ $b\bar b W^+$, where $h_{\rm SM}$ is the SM-like Higgs boson discovered at CERN in 2012 (either $h$ or $H$ in 2HDMs). The knowledge of its mass now provides in fact an additional handle in the kinematic selection when reconstructing a Breit-Wigner resonance in the $h_{\rm SM}\to b\bar b$  decay channel, thereby significantly improving the signal-to-background ratio afforded by pre-Higgs-discovery analyses \cite{whroy,me}. Such a study found  that, while this channel does not show much promise for a supersymmetric $H^\pm$   state, significant portions of the parameter spaces of several 2HDMs are testable at Run 2. 

Spurred by the aforementioned experimental results and building on Ref.~\cite{Uppsala}, we intend to study here all intermediate decay channels of a heavy $H^\pm$ state yielding a $b\bar b W^\pm$ signature, i.e., $H^+\to t\bar b$, $hW^\pm, HW^\pm$ and $AW^\pm$, starting from the production mode $bg\to tH^-$ + c.c. In doing so, we take into account interference effects between these four channels, in the calculation of the total $H^\pm$ width as well as of the total yield in the cumulative $b\bar b W^\pm$ final state (wherein the $W^\pm$ decays leptonically), with the aim of maximising the experimental sensitivity of ATLAS and CMS across them all. As intimated, we consider a 2HDM-II, as this uniquely predicts, amongst the various types, a heavy charged Higgs boson mass for consistency with flavour data.

The plan of this paper is as follows. In the next section we introduce the model and define its available parameter space based on current experimental and theoretical constraints. Then we proceed to a signal-to-background analysis. Finally, we draw our conclusions based on the results obtained. 

\section{Model implementation and parameter space constraints}

In order to discuss searches for the heavy $H^\pm$ state yielding a $b\bar b W^\pm$ signature we use as a benchmark model the 2HDM-II where the charged Higgs mass is constrained to be above 480 GeV due to $B$-physics constraints. Also, since we want to maximise the number of intermediate states that lead to $b\bar b W^\pm$, we choose the heavy CP-even scalar to be the SM-like 125 GeV one. This way, and by considering a light pseudoscalar, the decays $H^\pm \to A/h/H \, W^\pm \to b\bar b W^\pm$ are all possible, together with the $tb$ intermediate state. For such a heavy charged Higgs boson (with a mass above the top-quark one), the main production mechanism is $pp \to tH^-X + \mbox{charge conjugate}$, which depends strongly on $\tan \beta$. As we will discuss, the constraints from Run 1 force this scenario to be allowed only for low $\tan \beta$ values.

\begin{figure}[h!]
\begin{center}
\includegraphics[scale=0.75]{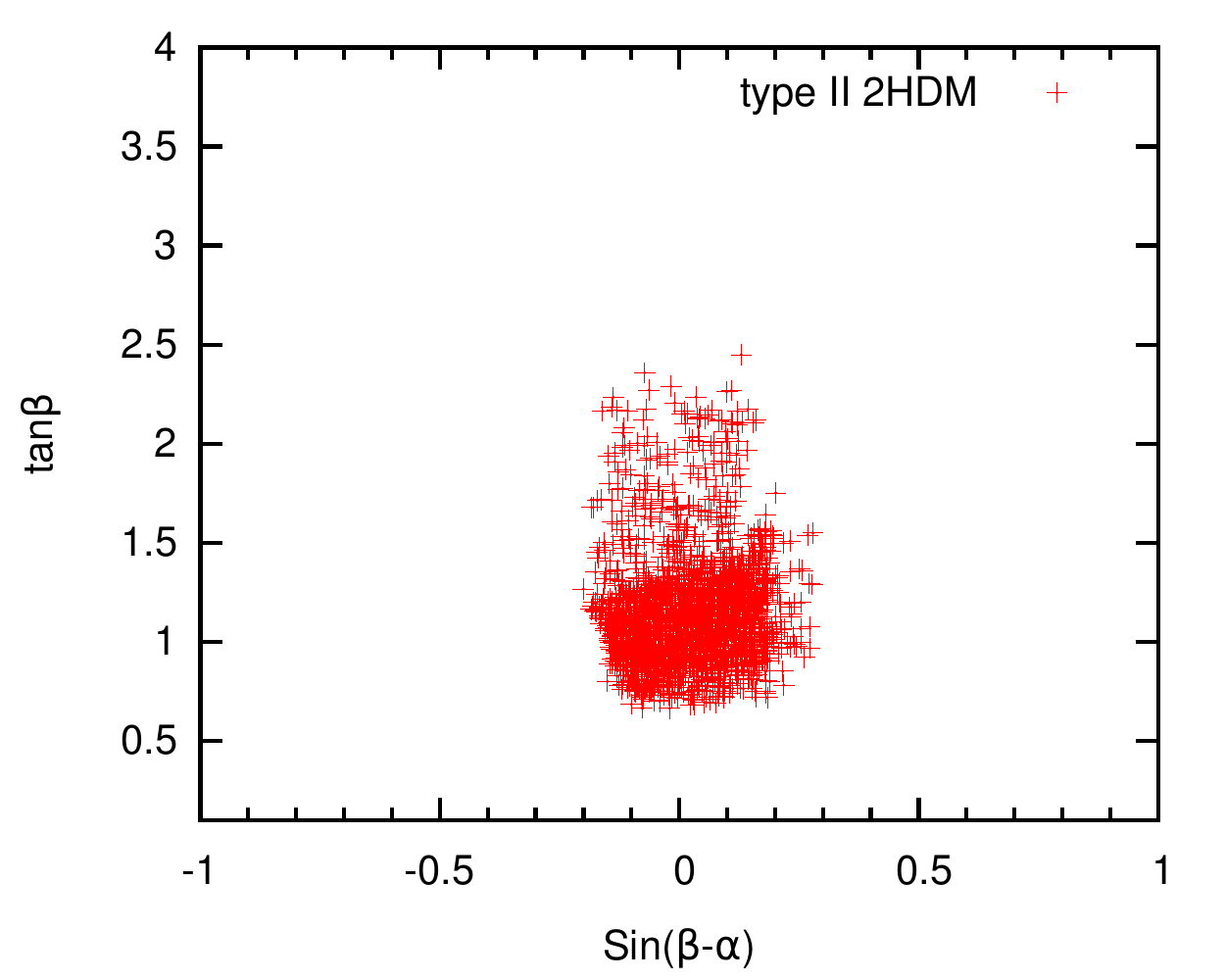}
\caption{\label{fig:points_2hdm} Allowed points on the $\sin(\beta-\alpha)$ versus $\tan\beta$ plane after the LHC Run 1.}
\end{center}
\end{figure}

We consider as free parameters of the model the four masses, $\tan \beta$, $\sin (\beta-\alpha)$, where $\alpha$ is the rotation angle in the CP-even sector, and the soft breaking parameter $m_{12}^2$. In the configurations yielding $M_H = 125$ GeV, 2HDMs, together with all theoretical and experimental constraints, were recently studied in detail in~\cite{Ferreira:2014dya}. The parameter space of 2HDMs is already very constrained by the LHC results obtained during the 7 and 8 TeV runs. In figure~\ref{fig:points_2hdm} we show the allowed parameter space at 95\% Confidence Level (CL) on the $(\tan\beta, \sin(\beta-\alpha))$ plane for Type-II with  $M_{H^\pm} = 500$ GeV, $M_H = 125$ GeV, $M_h = 80$ GeV and $M_A = 130$ GeV. We varied $m_{12}^2$ in its allowed range taking into account experimental and theoretical constraints as described in the \textsc{ScannerS} version for  CP-conserving 2HDMs~\cite{Ferreira:2014dya}. \textsc{ScannerS}~\cite{Coimbra:2013qq} is then interfaced with 2HDMC~\cite{Eriksson:2009ws} to obtain the Higgs decay rates. \textsc{HiggsBounds}\cite{Bechtle:2013wla} and \textsc{HiggsSignals}~\cite{Bechtle:2013xfa} are used to account for all collider results including the LHC ones. The value $\sin (\beta -\alpha) = 0$ corresponds to the case where the heavy Higgs has exactly the SM couplings to fermions and gauge bosons. When choosing our benchmark point it is clear from figure ~\ref{fig:points_2hdm} that $\tan \beta$ has to be small (but above 1) while $|\sin (\beta - \alpha)| \lessapprox 0.2$. Therefore the results will be presented for $\tan \beta =1$ and $\sin (\beta-\alpha) = 0.1$. Slightly larger values of $\tan \beta$ are allowed but because the cross section decreases with  $\tan \beta$ the sensitivity will become poorer as $\tan \beta$ increases. 
 
\section{Signal and background}

\begin{figure}[t!]
\begin{center}
\includegraphics[scale=0.65]{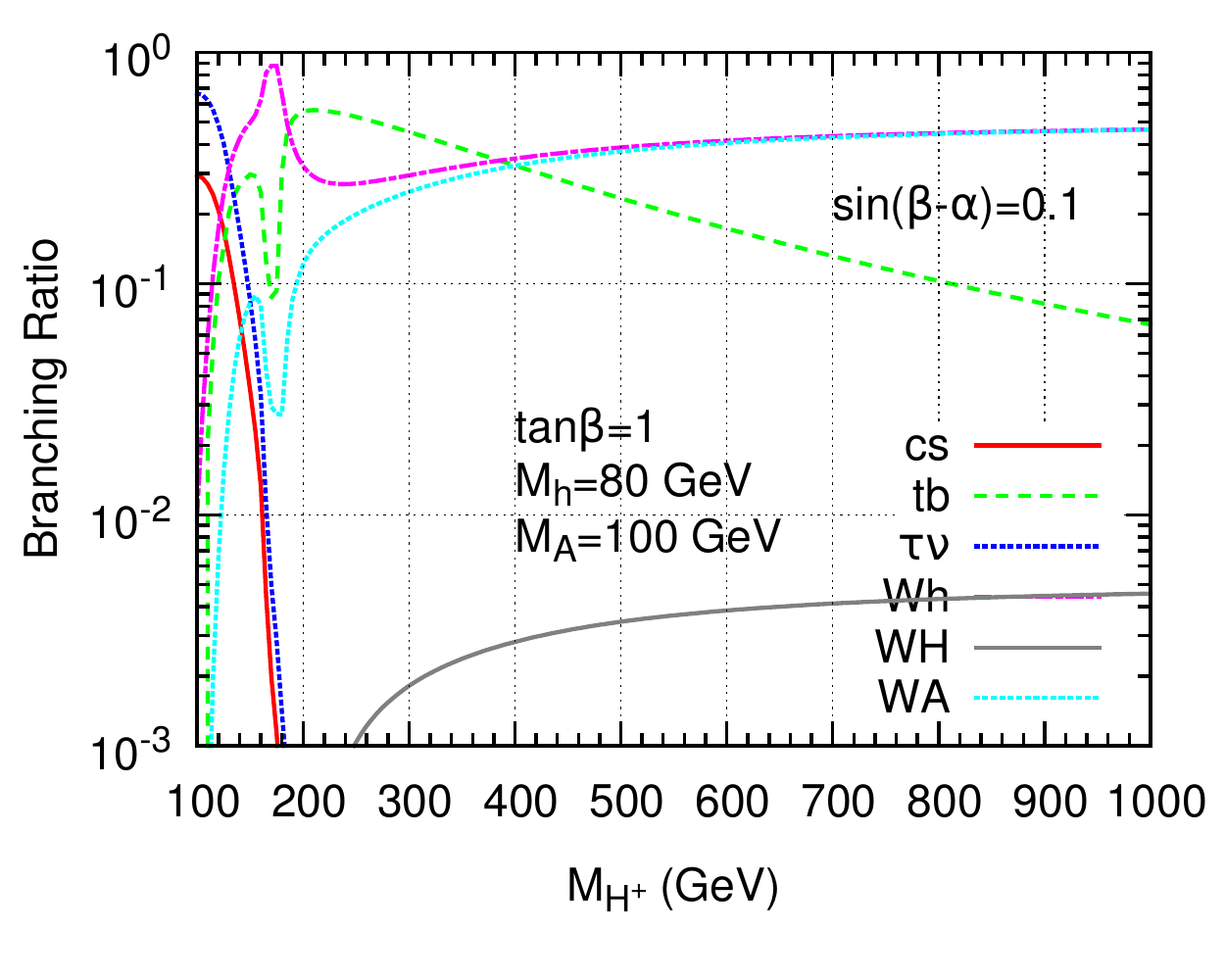}
\includegraphics[scale=0.65]{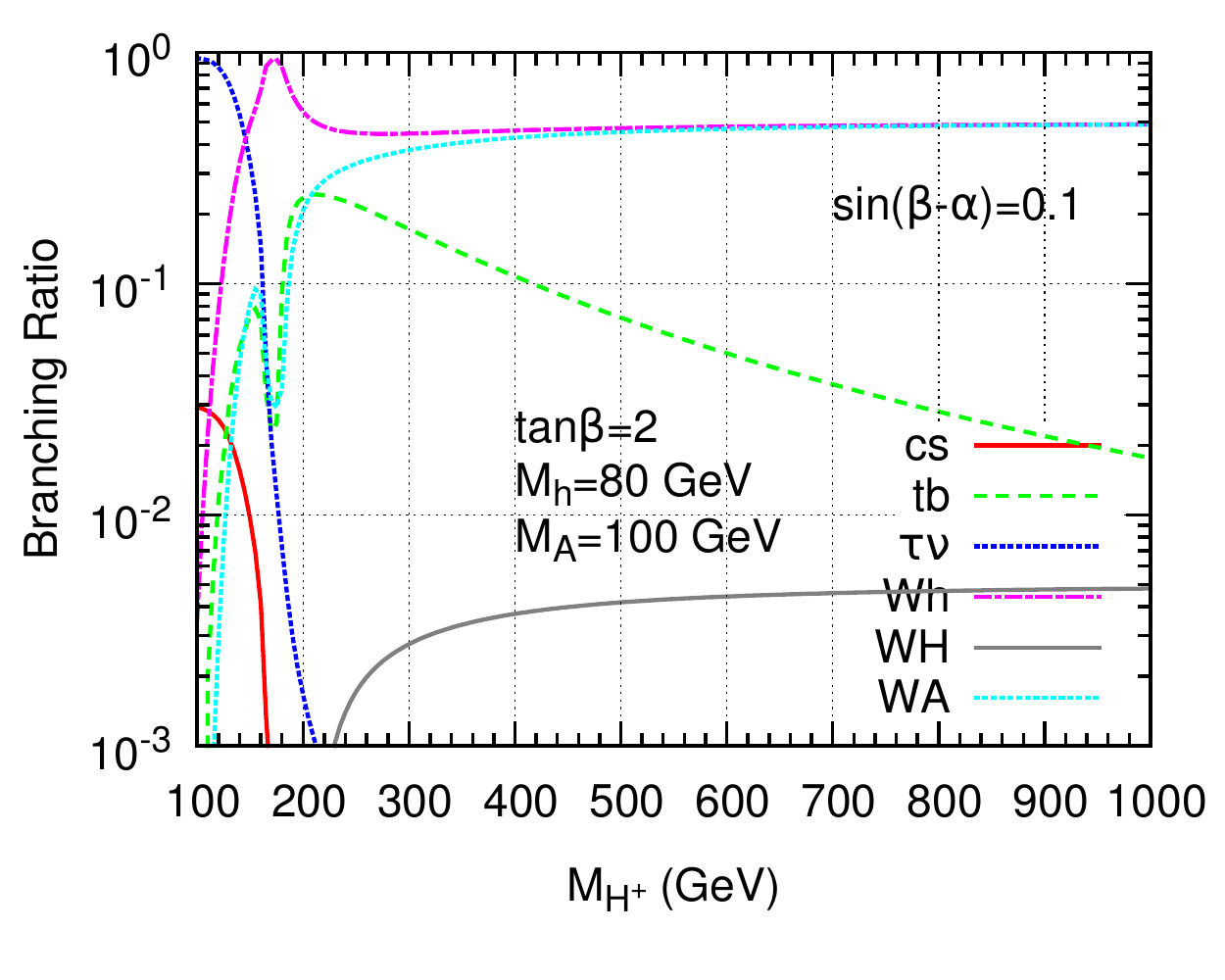}
\includegraphics[scale=0.65]{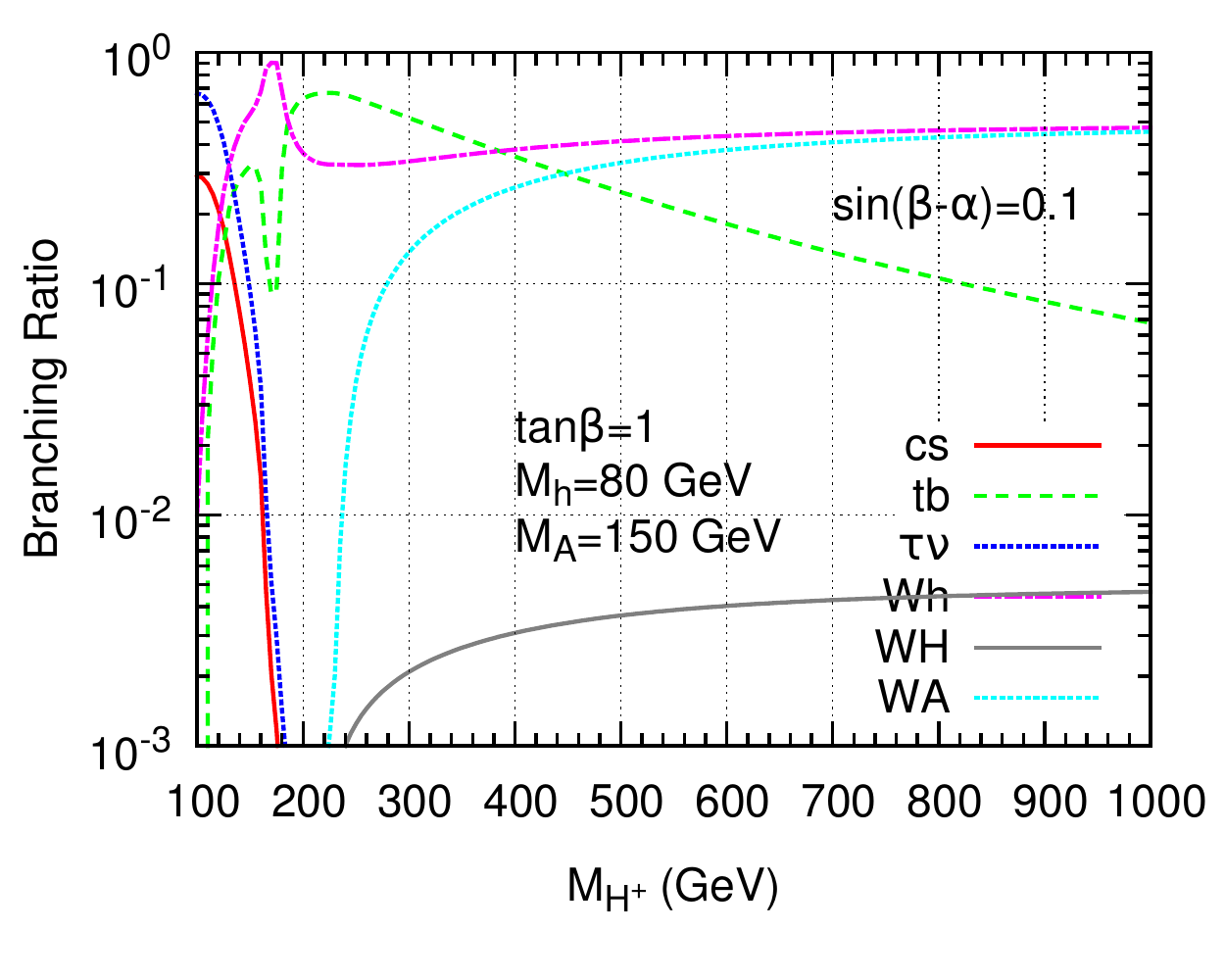}
\includegraphics[scale=0.65]{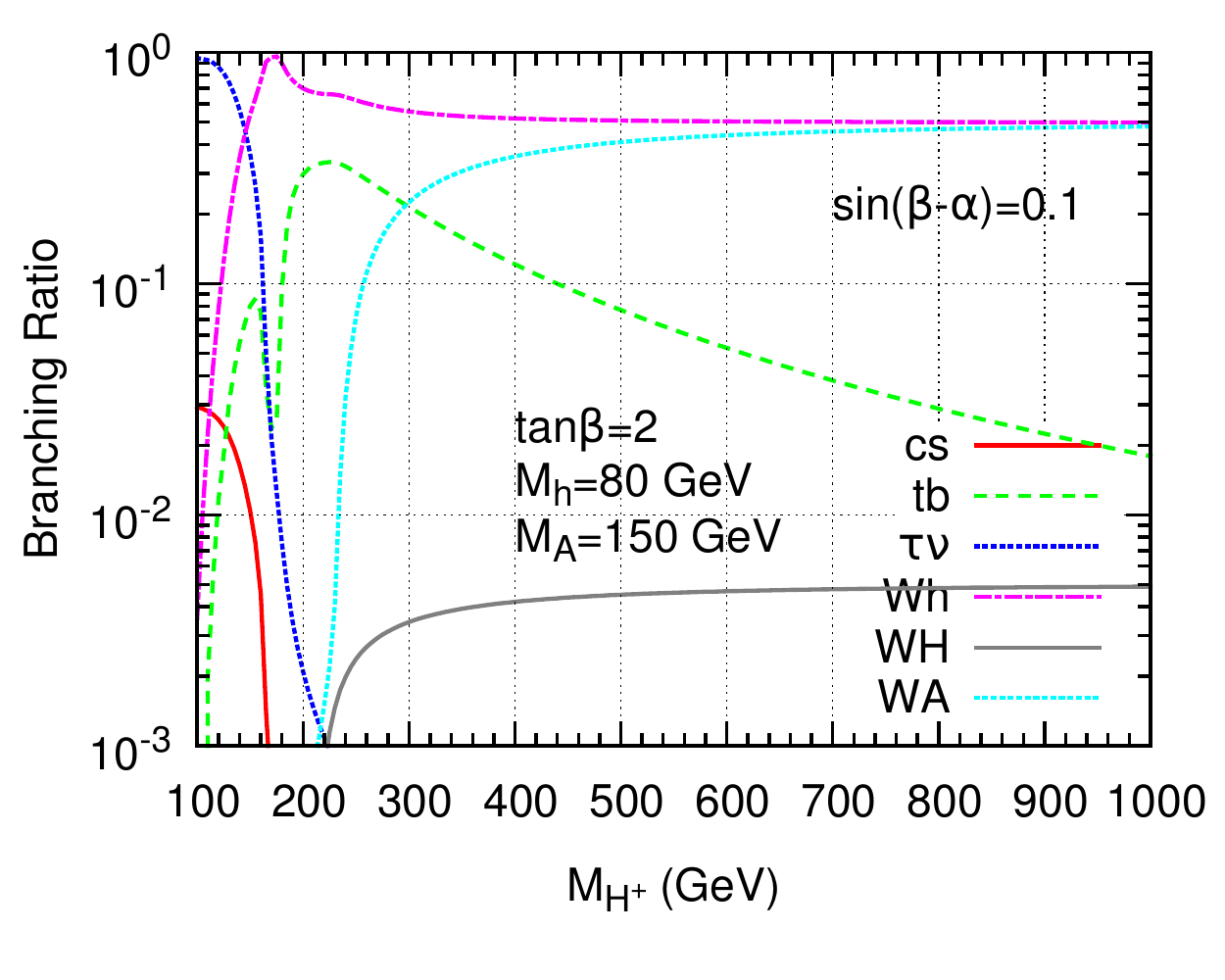}
\caption{\label{fig:br_2hdmII} BRs for a 2HDM-II charged Higgs boson with $M_{H^\pm} = 500$ GeV,  $\tan\beta=1$ and $M_A = 100$ GeV (upper left), $\tan\beta=2$ and $M_A = 100$ GeV (upper right), $\tan\beta=1$ and $M_A = 150$ GeV (lower left), $\tan\beta=2$ and $M_A = 150$ GeV (lower right). The remaining parameters are fixed to $M_H = 125$ GeV, $M_h = 80$ GeV and $\sin (\beta-\alpha) = 0.1$.}
\end{center}
\end{figure}

In the previous section we discussed the choice of a specific benchmark point for the 2HDM-II. The signal will thus be calculated for $M_{H^\pm} = 500$ GeV, $M_H = 125$ GeV, $M_h = 80$ GeV, $M_A = 130$ GeV $\tan \beta =1$ and $\sin (\beta-\alpha) = 0.1$ (the value of $m_{12}^2$ is not relevant because Higgs self-interactions do not actually take part in the processes
considered here). For this set of parameters, the charged Higgs boson has several decay channels kinematically allowed and with sizeable Branching Ratios (BRs). In figure~\ref{fig:br_2hdmII} we present all possible $H^\pm$ decay BRs for various $\tan\beta$ and $M_A$ choices. Clearly, because the scalars masses were chosen such that $M_{H^\pm} > M_{W^\pm}+M_X$ ($X$ being $h, H$ or $ A$), the decays
$H^\pm\to W^\pm X$ decays
 dominate over the $tb$ mode for a large portion of the parameter space. We recall here that in  2HDMs the charged Higgs couplings to a neutral scalar and a $W^\pm$ boson are 
\begin{eqnarray}
 & \Gamma^\mu_{H^\pm h W^\mp}& = \frac{g\cos(\beta-\alpha)}{2}(p_h-p_{H^\pm})^\mu,\\
 & \Gamma^\mu_{H^\pm H W^\mp}& = \frac{g\sin(\beta-\alpha)}{2}(p_H-p_{H^\pm})^\mu,\\
 & \Gamma^\mu_{H^\pm A W^\mp}& = \frac{g}{2}(p_A-p_{H^\pm})^\mu,
\end{eqnarray}
while couplings to quarks are Yukawa type dependent although they depend only on $\tan \beta$. If $h$ is the SM-like Higgs, then the LHC data forces $\sin(\beta-\alpha)\sim 1$ while if $H$ is SM-like then $\cos(\beta-\alpha)\sim 1$. Thus, in both cases, the charged Higgs boson couples more strongly to non-SM-like Higgs states than to the SM-like one. Unfortunately, this means that it is not possible to effectively search for $H^\pm$ in the $W^\pm b\bar b$ mode via the intermediate SM-like Higgs boson exploiting its invariant mass reconstruction around 125 GeV. As we have chosen $H$ to be the SM-like Higgs, a very heavy $H^\pm$ has the $W^\pm h$ and $W^\pm A$ modes as the dominant decays when compared to the $tb$  one. For the range of masses chosen for the scalars and $\tan \beta =1$, we obtain BR$(W^\pm A)$ $\sim$ BR$(W^\pm h) \sim 30-40\%$, BR$(W^\pm H)$ well below $1 \%$ and BR$(tb)\sim 20\%$ for $M_{H^\pm}=500$ GeV, while for $M_{H^\pm}=1$ TeV one has BR$(W^\pm A)$ $\sim$ BR$(W^\pm h)\sim 50\%$,  BR$(W^\pm H)$ well below $1 \%$ and BR$(tb)$ below $10 \%$. When $\tan\beta=2$, the decay width of the $tb$ mode decreases by a factor of 4 which leads to a further increase in the BR of the $W^\pm h$ and $W^\pm A$ modes. Thus, for a very heavy $H^\pm$ and larger $\tan\beta$, the $tb$ decay mode becomes negligible whenever the remaining scalars are light.


The main production mode for a heavy charged Higgs boson is the associated process $(pp\to tH^-X + \mbox{charge conjugate})$. The production cross section is $\sim 900$ fb (including charge conjugate cross section) for $M_{H^\pm}=500$ GeV and $\tan\beta=1$ in 2HDM-II at Leading Order (LO) in QCD.
%
All four decay modes of a heavy charged Higgs lead to the $W^\pm \bar b b$ final state when all neutral scalars decay to a $b\bar b$ pair and $t\to b W^+$. Our final signal will 
therefore be $H^\pm$ production in association with a top-quark followed by the decay of $H^\pm \to W^\pm b\bar b$ and $t\to b W^\pm$. Hence, the signal contains two $W^\pm$ bosons and 3 $b$-tagged jets. We will consider the case where one of the $W^\pm$ bosons decays leptonically while the other decays into a pair of light jets. Therefore, the final signal process contains at least one lepton, at least 2 light jets, at least 3 $b$-jets and missing transverse energy.

The only irreducible background to the signal comes from the $WWbbb$ process with a LO cross section of about 10 pb at the 14 TeV LHC. The $WWbbb$ background includes its main contribution which originates from $t\bar tb$. Other single top backgrounds like $tW^- h$, $tW^- Z$, $tW^- A$ and $tW^- g$ with $h/A/Z/g \to b\bar b$ are also included in the $WWbbb$ contribution which we consider in the analysis. 
There are however QCD backgrounds resulting from  light-quark and gluon jets faking the $b$-jets. The largest non-irreducible background is $WWbbj$ when the light jet is misidentified as a $b$-jet. The dominant contribution to the $WWbbj$ background has its origin in the $t\bar t j$ process. Finally, we have also considered the $WWbjj$ noise when two light-quark and/or gluon jets are misidentified as two $b$-jets. This background is however very small after all cuts are taken into account, amounting to about 3 to 7 \% of the total background.

The signal and background events were generated at LO using {\tt Madgraph5}~\cite{madgraph}. Further, we have used {\tt PYTHIA8.2}~\cite{pythia8} to perform parton shower and hadronisation for both signal and background events. Then we carried out a full detector simulation with {\tt DELPHES3}~\cite{delphes}, which is a framework for the fast 
emulation  of a generic collider experiment. For detector and trigger configurations, we resorted to the ATLAS default card. 

\subsection{Selection}

We now describe the selection and analysis cuts in detail\footnote{In all upcoming figures the observables will all be normalised to the same area.}. All the events should satisfy the following selection and identification cuts.

\begin{figure}[t!]
\begin{center}
 \includegraphics[scale=0.35]{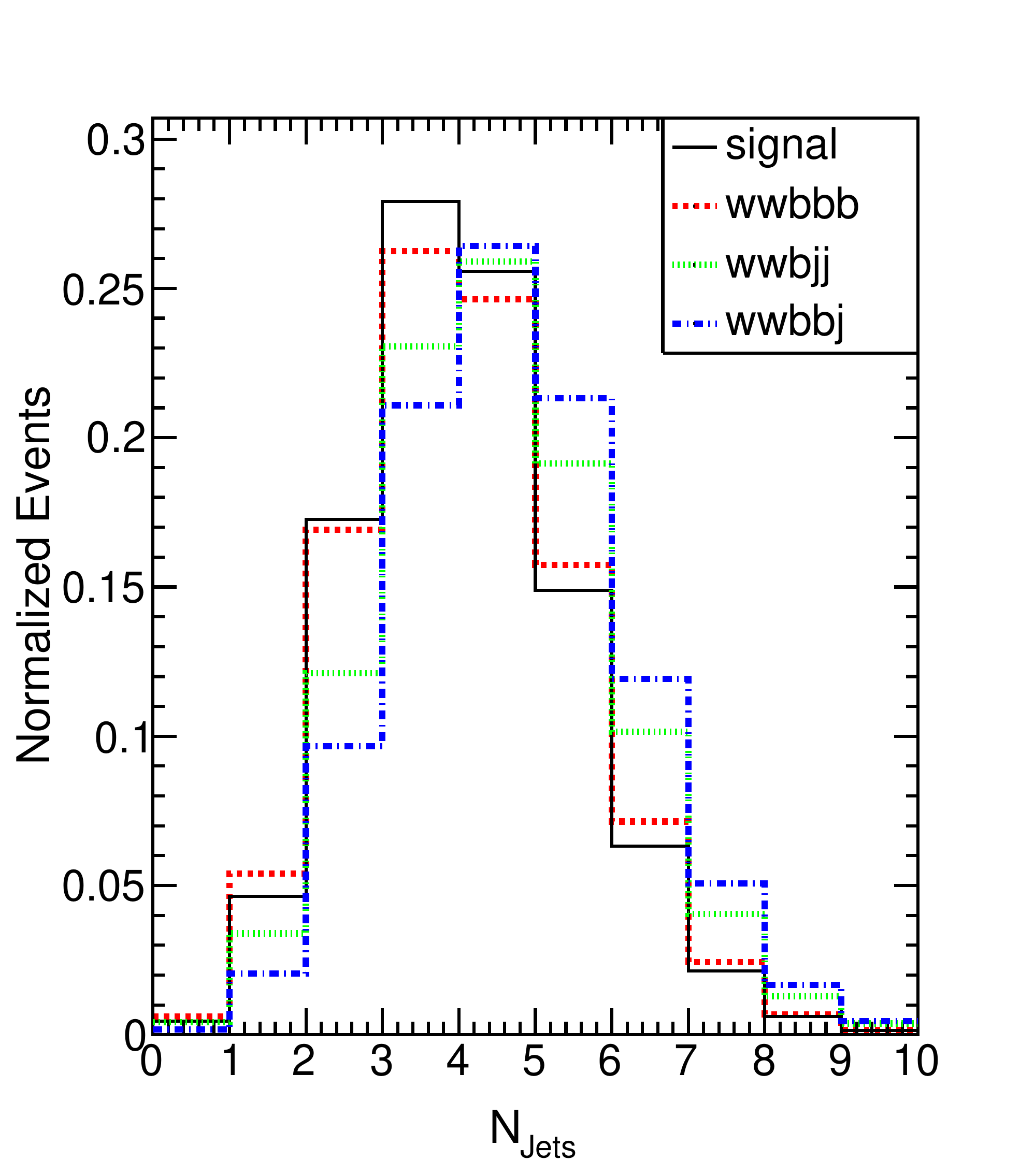}
 \includegraphics[scale=0.35]{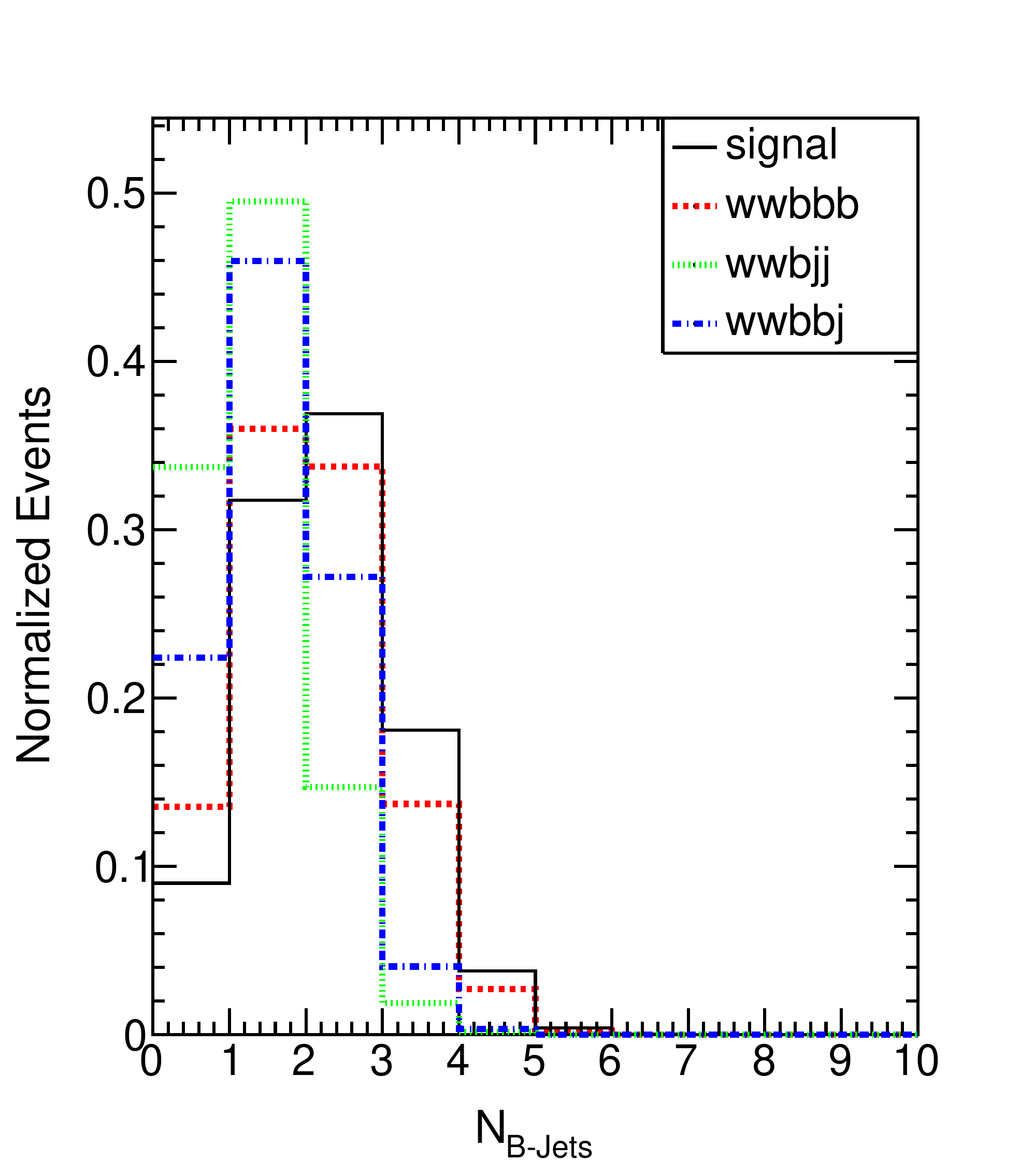}
\caption{\label{fig:nbjet} Number of light jets (left) and $b$-tagged jets (right) in an event for the signal and backgrounds.}
\end{center}
\end{figure}

\begin{itemize}
\item {\bf Identification cuts}
 \begin{enumerate}
  \item Events must have at least one lepton ($e$ or $\mu$), 3 $b$-jets and at least 2 light jets.
  \item Leptons must have transverse momentum $p_T>20$ GeV and rapidity $|\eta|<2.5$.
  \item All jets must satisfy the following $p_T$ and $\eta$ requirements:
  $$p_{Tj}>20~ \mbox{GeV}, |\eta_j|<2.5.$$
  \item All pairs of objects must be well separated from each other,
  $$\Delta R_{jj,jb,bb,\ell j,\ell b}\geq 0.4~~ \mbox{where}~~\Delta R=\sqrt{(\Delta \phi)^2+(\Delta \eta)^2}.$$
 \end{enumerate}

\item{\bf Efficiency for $b$-jet (mis-)identification}

For  $b$-tagging, we use the improved value of the efficiency from the ATLAS new $b$-tagging algorithm \cite{btag}. That is, in this analysis, we use a $b$-tagging efficiency according to following rule: $$\epsilon_{\eta}\tanh(0.03\;p_T-0.4),$$ where $\epsilon_\eta=0.7$ for $|\eta|\leq 1.2$ and 0.6 for $1.2\leq|\eta|\leq 2.5$. We use this same expression for the probability of a $c$-jet faking a $b$-jet but now with $\epsilon_\eta=0.2$ for $|\eta|\leq 1.2$ and $\epsilon_\eta=0.1$ for $1.2\leq|\eta|\leq 2.5$. Finally, for the light-quark and gluon jets, we take the mistagging probability to be 0.001 throughout. In figure~\ref{fig:nbjet} we show the number of $b$-tagged jets for the signal and backgrounds. As we require at least 3 $b$-tagged jets in our analysis, this requirement alone reduces the $WWbbj$ background events by a factor of $10^{3}$.

\item {\bf Selection requirements} 

When an event satisfies all above requirements, it is further processed for signal reconstruction and background reduction as follows.

\begin{enumerate}
\item {\bf Cut on $H_T$}:  a useful variable is the scalar sum of the $p_T$'s of all the visible particles in the final state, 
\begin{equation}
 H_T=p_T^{\ell^\pm}+\sum_j p_T^j.
\end{equation}
Figure~\ref{fig:HT} shows the $H_T$ distributions for the signal and backgrounds. In the figure we see that the peak of the scalar $H_T$ distribution for the signal is around 600 GeV while for the backgrounds it is around 400 GeV. This is due to the fact that the signal events include a heavy particle which produces high-$p_T$ decay products. A cut on $H_T > 500$ GeV reduces the $WWbbj$ and $WWbbb$ backgrounds to 36\% and 27\% of their initial values, respectively, while the signal events are only decreased to 87\% of their initial values, as can be seen from table \ref{tab:14Tev_cutflow}. This cut plays therefore a crucial role in increasing the $S/B$ and $S/\sqrt{B}$ ratios. Quantitatively, the $H_T>500$ GeV cut increases the $S/\sqrt{B}$ significance of the signal from 3.8 to 6.1 for 300 fb$^{-1}$ of integrated luminosity.

\begin{figure}[h!]
\begin{center}
\includegraphics[scale=0.35]{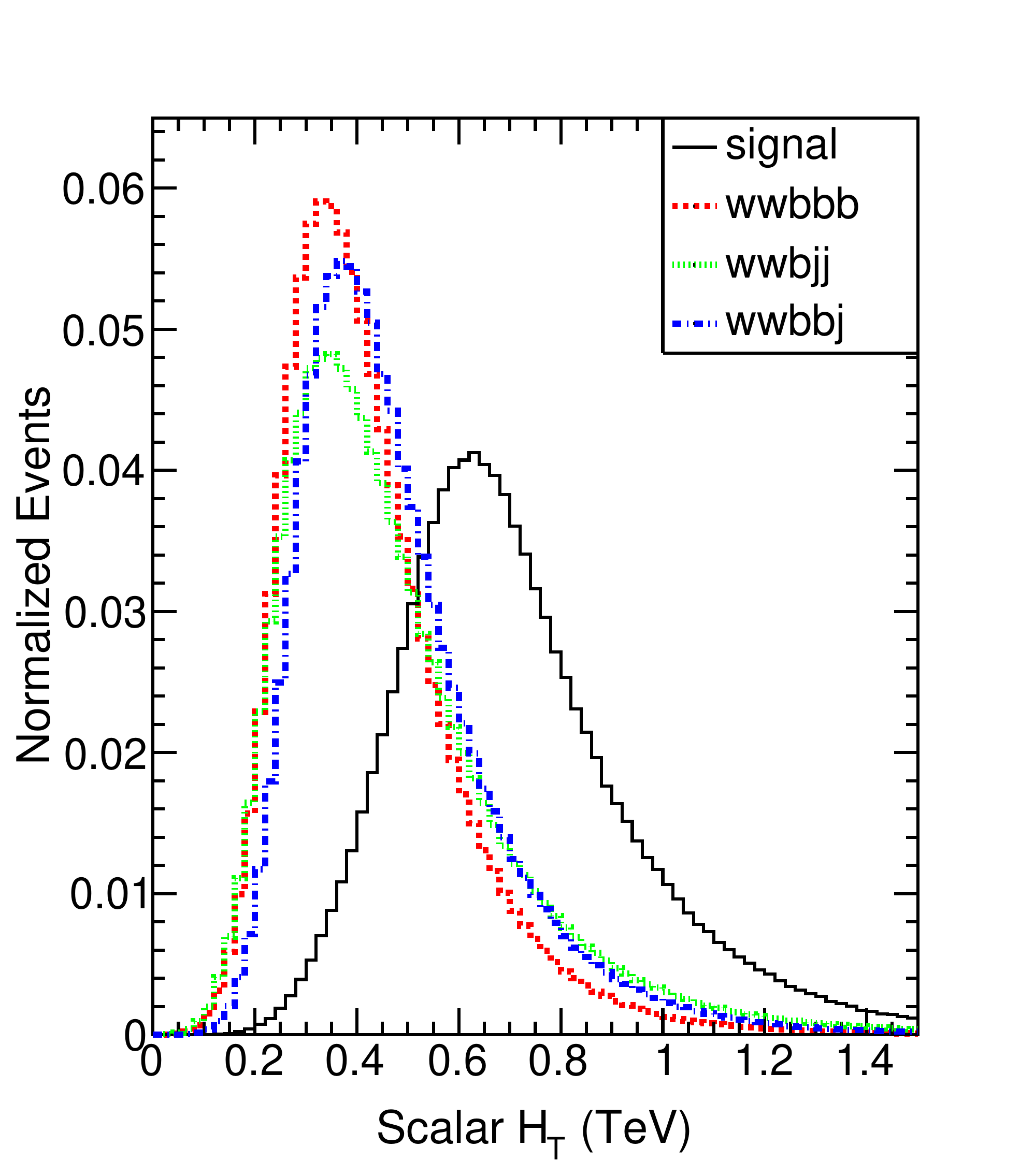}
\caption{\label{fig:HT} Scalar sum of $p_T$'s ($H_T$) distribution for signal and backgrounds.}
\end{center}
\end{figure}

\item {\bf $\Delta R$ separation}: in figure~\ref{fig:delr_jb} we show the $\Delta R$ separation between the hardest $b$-jet and the hardest light jet (left) and between the hardest $b$-jet and second hardest light jet (right) for the signal and backgrounds. As the charged Higgs boson is heavy, it is expected to be produced with low momentum. Thus, when it decays to, e.g., $h$ and $W^\pm$, they would be highly boosted with each moving in different hemispheres. When $h$ and $W^\pm$ further decay to pairs of $b$-jets and light jets, respectively, the separation between the $b$-jets and the light jets is expected to be considerably larger. This fact is confirmed by figure~\ref{fig:delr_jb}. As a consequence we observe a more pronounced peak for the signal at large $\Delta R$ for both the $j_1b_1$ and $j_2b_1$ combinations, where the label 1(2) refers to the hardest(second hardest) jet. Based on these observations, we put another cuts on $\Delta R$ to suppress the backgrounds: $\Delta R_{jb}>2.0$ since the noise  has a considerably larger number of events in the region $\Delta R_{jb}<2.0$.

\begin{figure}[h!]
\begin{center}
\includegraphics[scale=0.35]{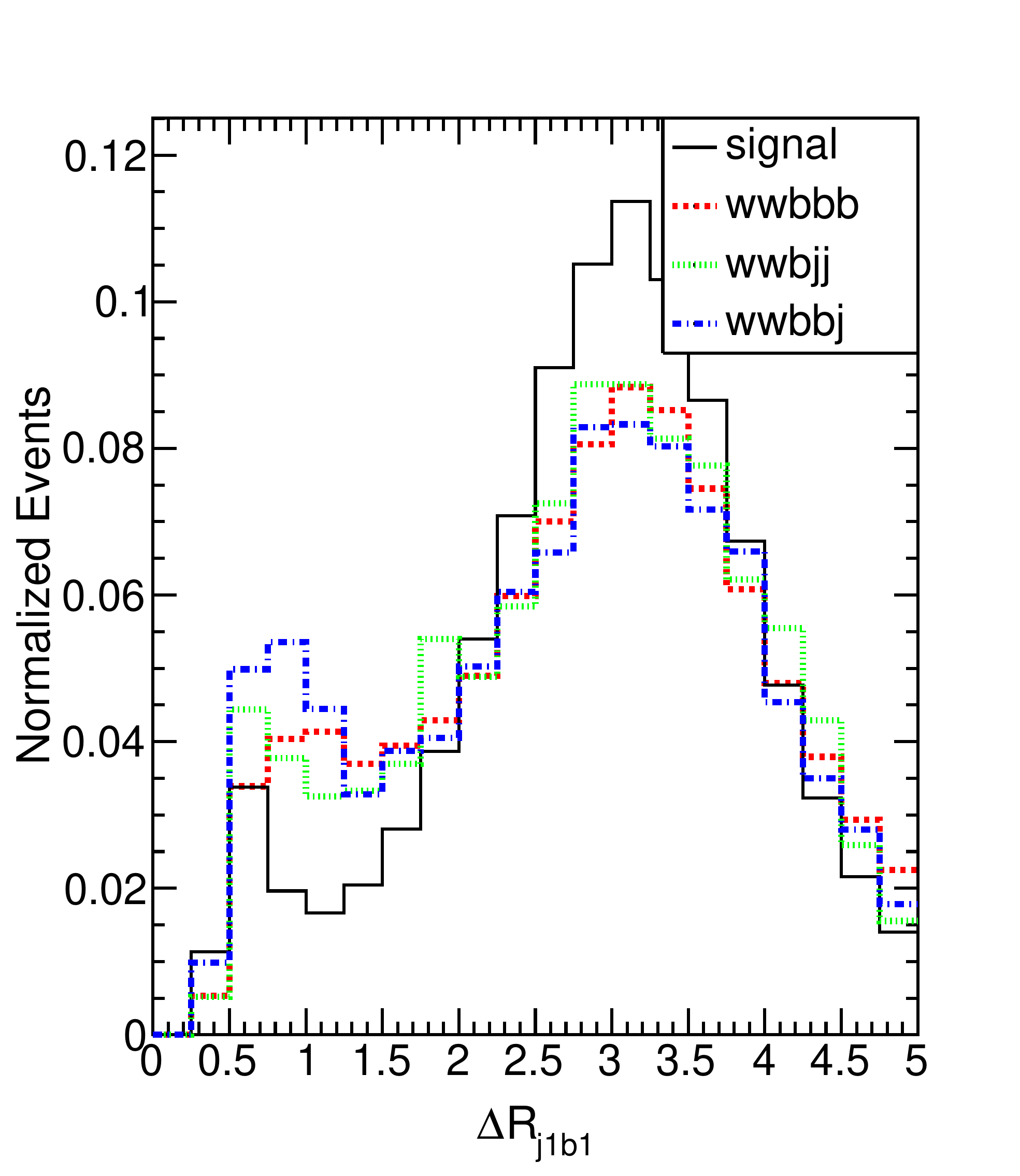}
\includegraphics[scale=0.35]{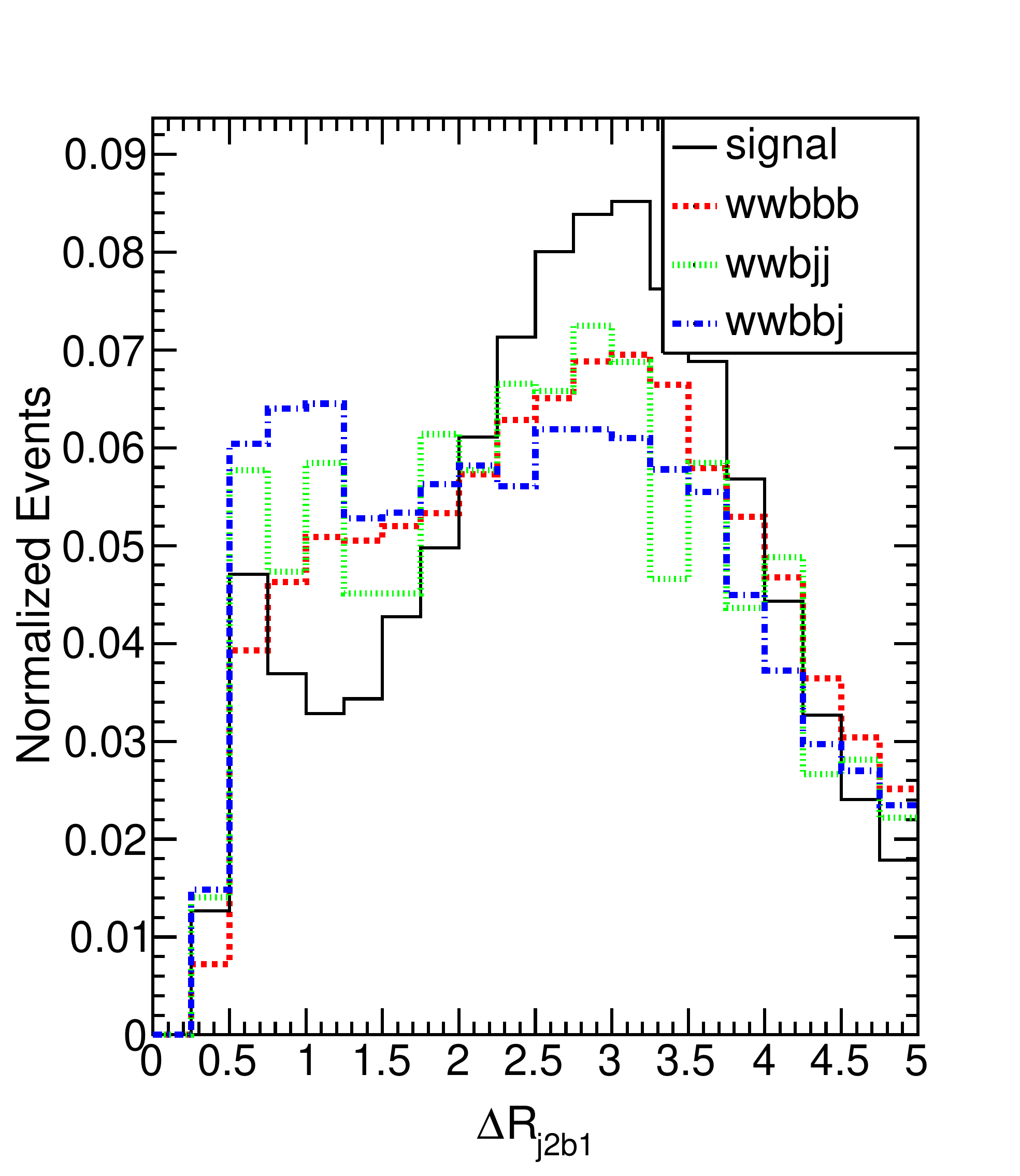}
\caption{\label{fig:delr_jb} $\Delta R$ separation between the hardest $b$-jet and the hardest light jet (left) and between the hardest $b$-jet and second hardest light jet (right) for the signal and backgrounds. }
\end{center}
\end{figure}

\item {\bf Hadronic $W^\pm$ candidate}: as we are considering a very heavy charged Higgs state, it is expected that its decay products, the $W^\pm$ and $b\bar b$ pairs, are highly-boosted. This in turn means that these bosons decay to closely spaced final states. 
We search for non-$b$ tagged jets and take the pair with minimum $\Delta R$ to form the hadronic $W^\pm$ candidate. In the left panel of figure~\ref{fig:delr_jj} we present the $\Delta R$ separation between the two hardest light-flavour jets. We find that the $\Delta R$ distribution for the two hardest light jets peaks at a very low value of $\Delta R$  and the jets are thus very closely spaced as expected. We reconstruct the jets with $\Delta R_{\mathrm{min}}$ to form the hadronic $W^{\pm}$. In the right panel of figure~\ref{fig:delr_jj}, we show the invariant mass distribution of the two light-flavour jets which have minimum $\Delta R$ separation. We find that the distribution peaks at the $W^\pm$ boson mass. Thus, to suppress the background, we further collect only events lying within a mass window of 25 GeV: {\it viz.} $$|M_{jj}-M_{W^\pm}|<25~ \mbox{GeV}.$$

\begin{figure}[h!]
\begin{center}
\includegraphics[scale=0.35]{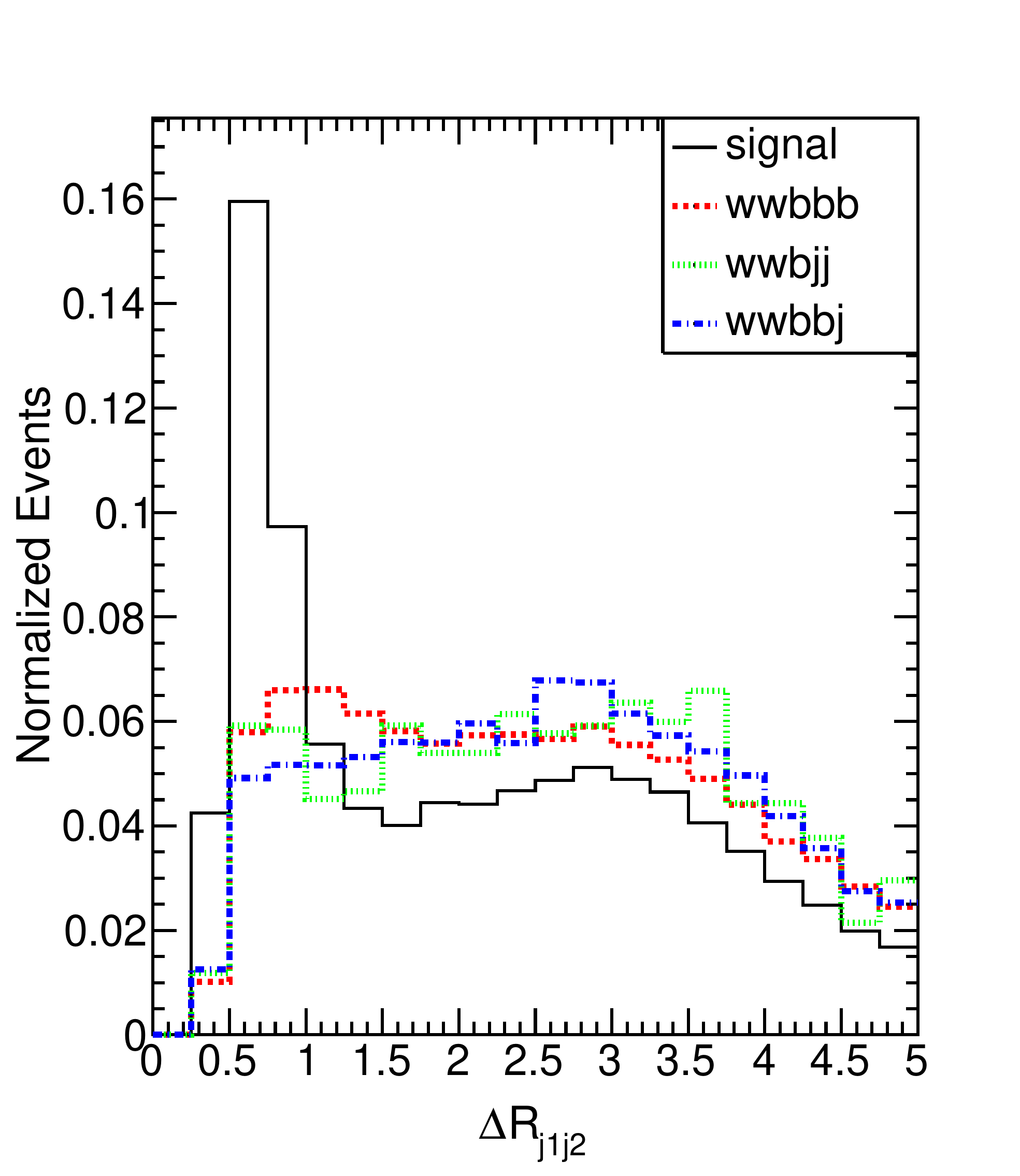}
\includegraphics[scale=0.35]{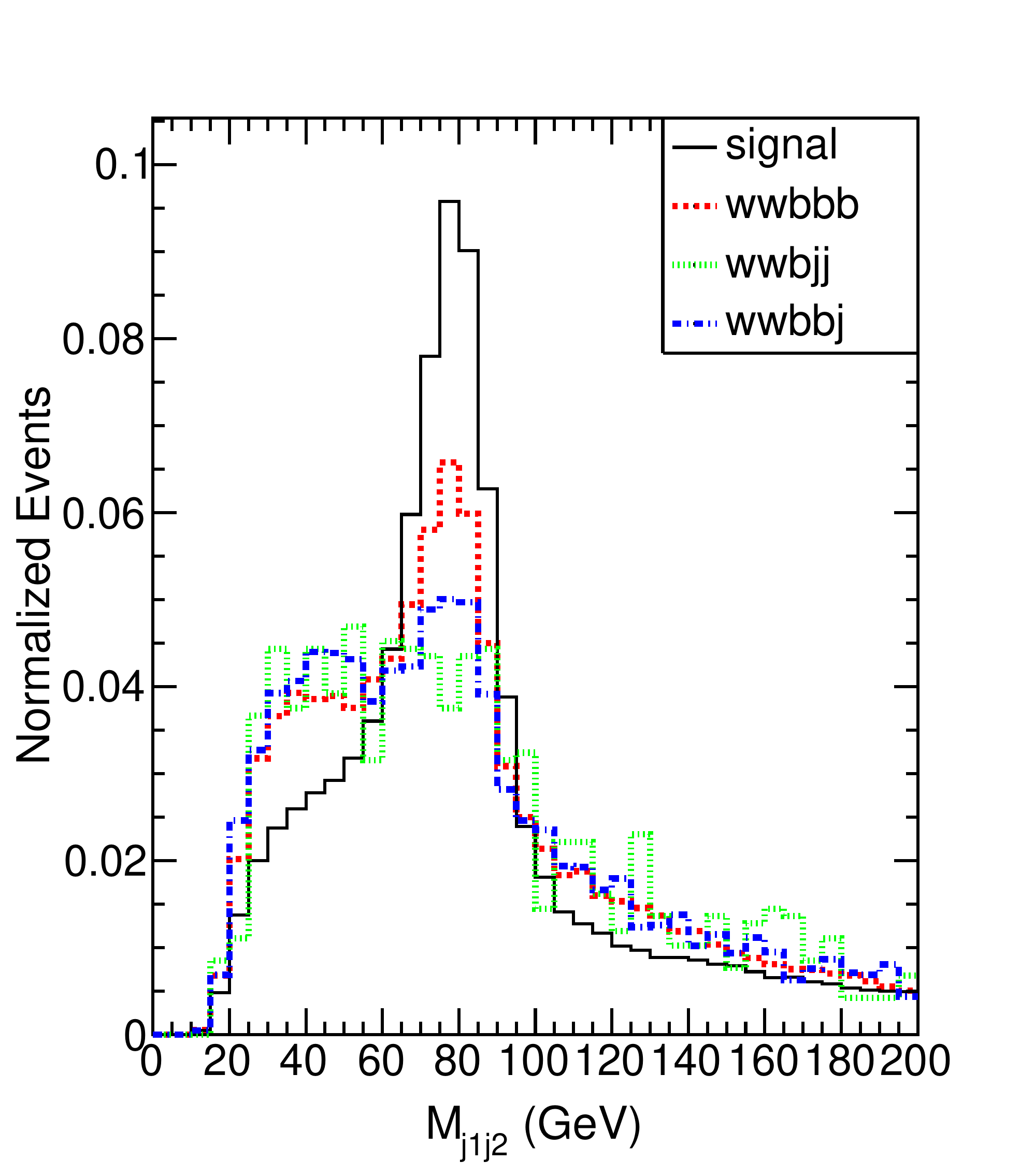}
\caption{\label{fig:delr_jj}$\Delta R$ separation between the two hardest light-flavour jets (left) and their invariant mass  $M_{j_1 j_2}$ (right) for the signal and backgrounds.}
\end{center}
\end{figure}

\begin{figure}[h!]
\begin{center}
\includegraphics[scale=0.35]{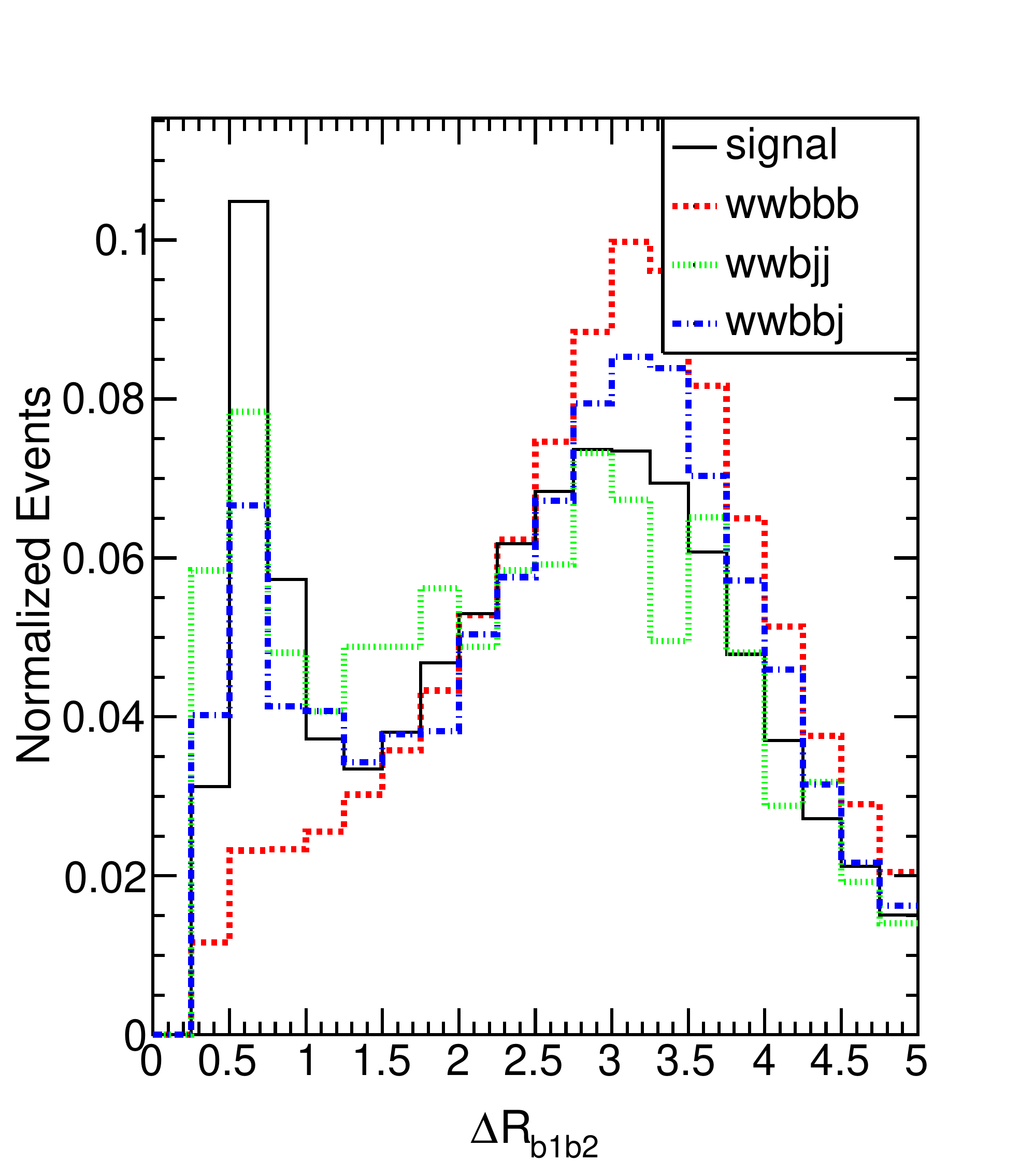}
\caption{\label{fig:delr_bb} $\Delta R$ separation between the two hardest $b$-jets for the signal and backgrounds. }
\end{center}
\end{figure}

\item {\bf Leptonic $W^\pm$:}  the momentum of the neutrino coming from the leptonically decaying $W^\pm$ is determined using the information about the missing transverse momentum. Imposing the invariant mass constraint $M_{l\nu}^2 = M_{W^\pm}^2$, we obtain the longitudinal component of the neutrino as
\begin{equation}
 p_{\nu L}=\frac{1}{2p_{\ell T}^2}\left(A_W p_{\ell L} \pm E_\ell \sqrt{A_W^2\pm 4 p_{\ell T}^2 E_{\nu T}^2}\right),
\end{equation}
where $A_W=M_{W^\pm}^2+2p_T\cdot E_{\nu T}$. If two solutions for $p_{\nu L}$ are found, the one which gives $M_{l\nu}$ closer to the $W^\pm$ mass is adopted. Also, we reject the events with complex solutions. Using the momenta of the reconstructed neutrino and  lepton, the momentum of the leptonic $W^\pm$ can be obtained.

\begin{figure}[h!]
\begin{center}
\includegraphics[scale=0.3]{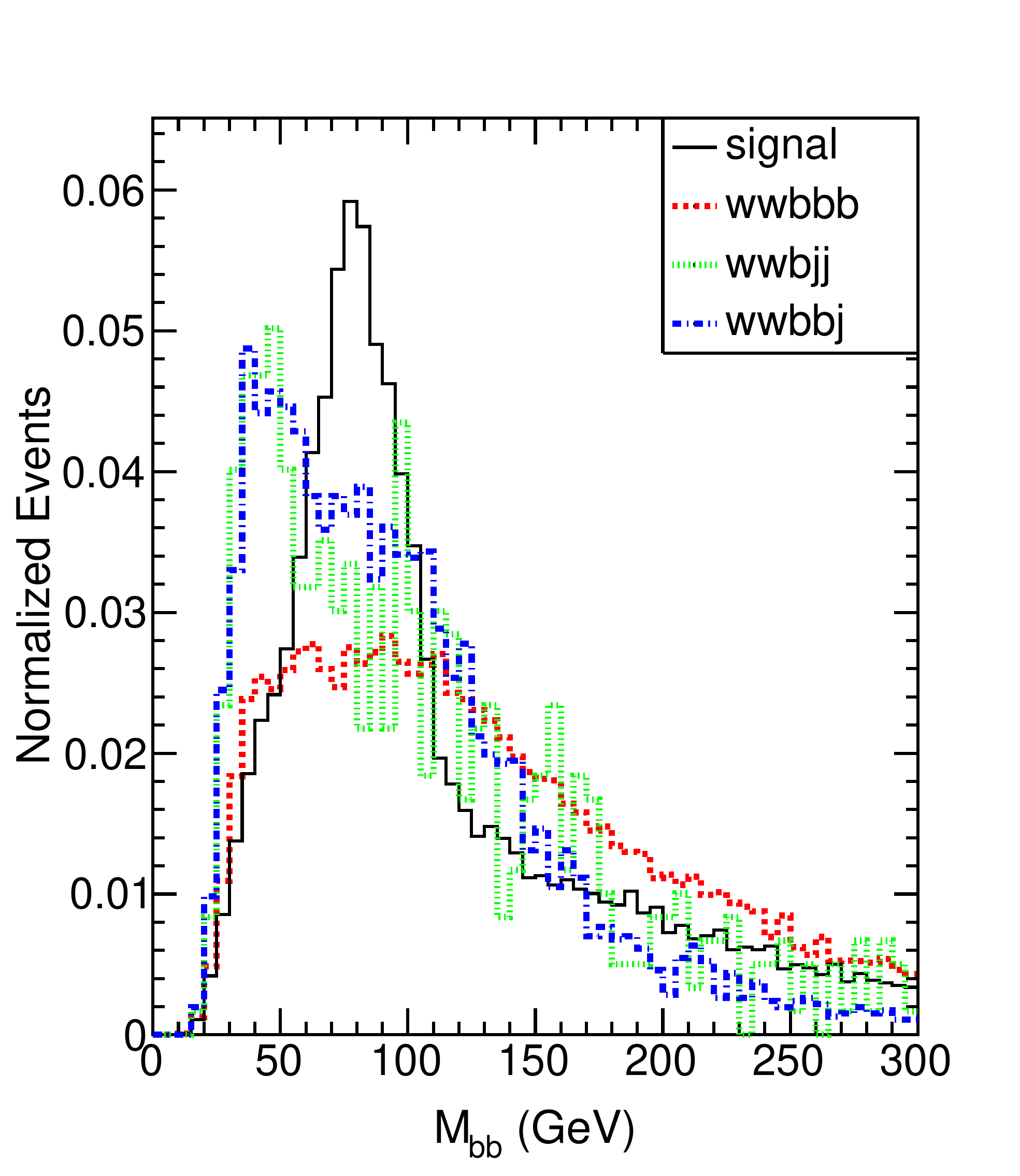}\hspace{-0.25cm}
\includegraphics[scale=0.3]{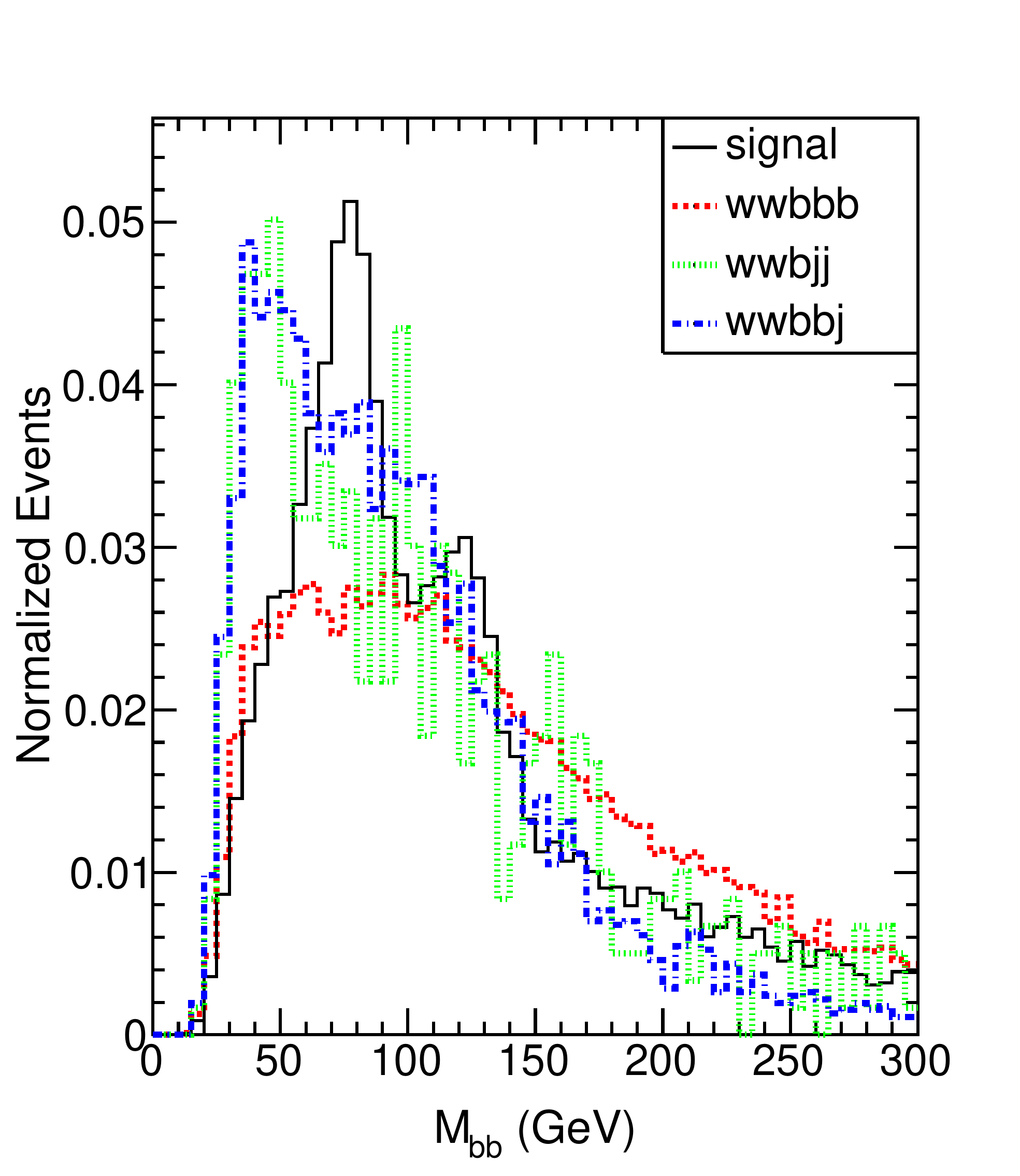}\hspace{-0.25cm}
\includegraphics[scale=0.3]{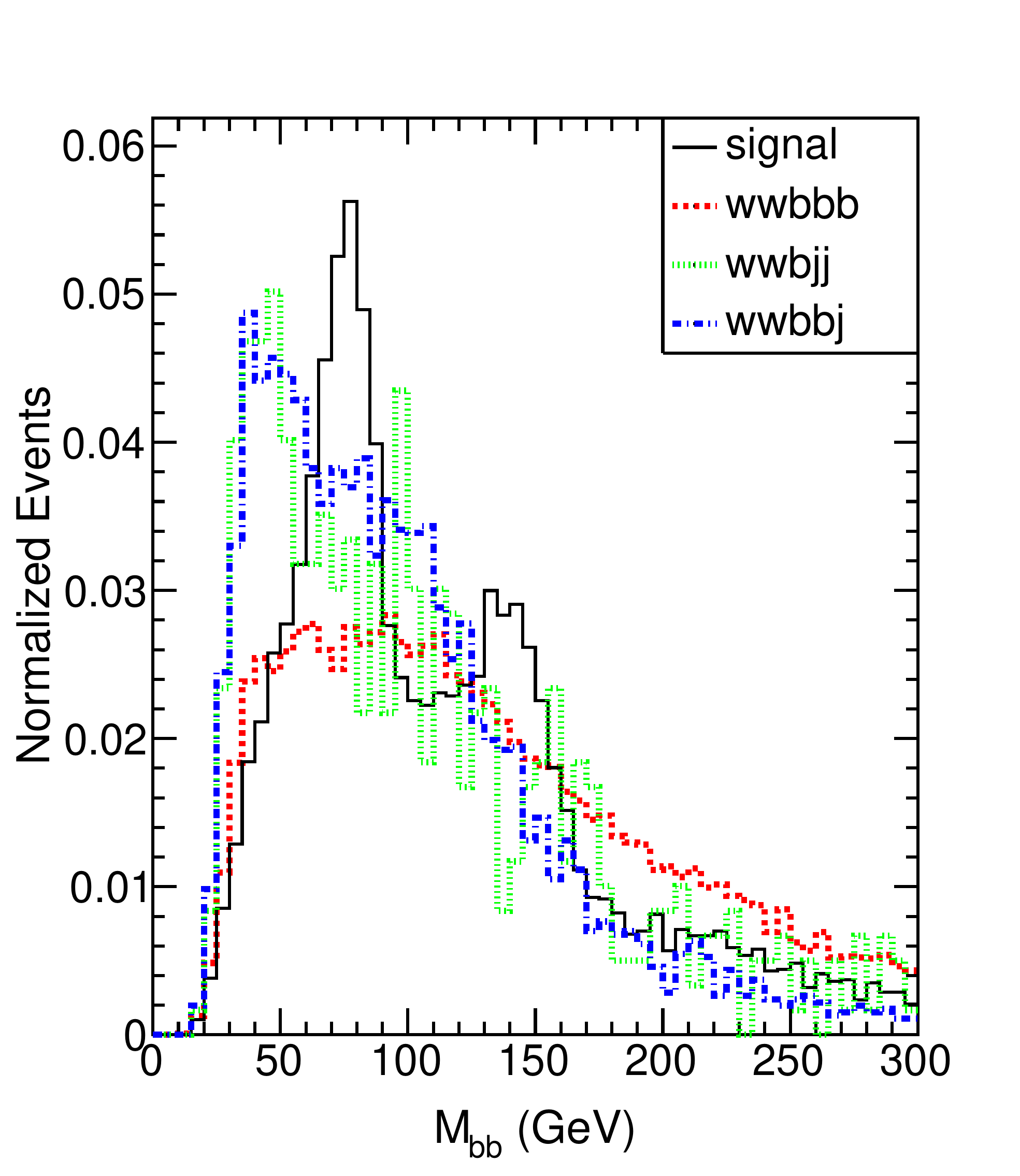}
\caption{\label{fig:INV_Mbb} Invariant mass ($M_{bb}$) of two $b$-jets with $\Delta R_{\rm min}$ for $M_A=$ 100 GeV (left), 130 GeV (middle) and 150 GeV (right)  for the signal and backgrounds. As discussed in the text, the two $b$ jets with  minimum $\Delta R$ are chosen to reconstruct $h,~A$.}
\end{center}
\end{figure}

\end{enumerate}

At this point of the analysis, we have reconstructed  $WW b b b$ events\footnote{Note that we do not enforce charge reconstruction of a $b$-tagged jet.}. In table~\ref{tab:14Tev_cutflow}, we show the corresponding cut flow of the cross section for the signal and individual backgrounds. Next we devise a dedicated set of cuts in order to extract our signals, {\it viz.}, $W^\pm h$, $W^\pm A$ and $tb$\footnote{Recall that the $W^\pm H$ (with the $H$ state being SM-like) component of the signal is negligible.} from the already reconstructed $WWbbb$ events. The extraction of each individual signal requires a customised set of cuts which we will now discuss.

\begin{table}[h!]
\begin{center}
{\renewcommand{\arraystretch}{1.5}}
 \newcolumntype{C}[1]{>{\centering\let\newline\\\arraybackslash\hspace{0pt}}m{#1}} 
\begin{tabular}{ ||C{0.85cm}C{2.5cm}C{1.75cm}|| C{1.5cm} | C{1.5cm} | C{1.5cm} |C{1.5cm} |C{2.25cm}||}
\hline \hline
\multicolumn{3}{||c||}{\multirow{2}{*}{Cuts}}&\multicolumn{5}{c||}{$\sigma$ [fb] } \\\cline{4-8}
&&&Signal & $WWbbj$  & $WWbbb$  & $WWbjj$ & Total Background   \\\cline{1-8}
C0:& \multicolumn{2}{c||}{No Cuts}& 235.0	& 1.6$\times 10^5$		& 5.2$\times 10^3$	& 3.3$\times 10^4$ & 2.0$\times 10^5$	\\ \hline
C1:& \multicolumn{2}{c||}{$H_T>500$ GeV }& 205.0	& 5.8$\times 10^4$		& 1.4$\times 10^3$	& 1.2$\times 10^4$ & 7.1$\times 10^4$	\\ \hline
\multirow{4}{*}{C2:} &$ \Delta R_{ij} > 0.4$ & $i,j=b,j,\ell$
& \multirow{4}{*}{203.1}& \multirow{4}{*}{5.7$\times 10^4$}& 		\multirow{4}{*}{1.3$\times 10^3$}& \multirow{4}{*}{1.1$\times 10^4$} &\multirow{4}{*}{7.0$\times 10^4$}\\
&  $p_{T}^b > 25 \ \mbox{GeV}, $     &$|\eta_b|<2.5$&&&&&\\
&  $p_{T}^\ell > 25 \ \mbox{GeV}$,    &$|\eta_\ell|<2.5$&&&&&\\
&  $ p_{T}^j > 25 \ \mbox{GeV}$,    &$|\eta_j|<2.5$&&&&&\\   
\hline
C3:& \multicolumn{2}{c||}{Only one lepton} 					& 102.4 & 3.3$\times 10^4$ 	& 714.0 & 2.6$\times 10^3$ & 3.7$\times 10^4$	\\ \hline
C4:& \multicolumn{2}{c||}{ At least 2 light jets} 				& 97.6	& 3.2$\times 10^4$ 	& 671.6	& 2.5$\times 10^3$ & 3.5$\times 10^4$	\\ \hline
C5:& \multicolumn{2}{c||}{At least 3 $b$-tagged jets}				& 34.4	& 1.3$\times 10^3$	& 100.9	& 44.8 & 1.5$\times 10^3$		\\ \hline
C6:& \multicolumn{2}{c||}{Cuts on $\Delta R_{j_1b_1}$ \& $\Delta R_{j_1b_2}$}	& 33.1	& 1.1$\times 10^3$	& 87.5	& 43.1 & 1.2$\times 10^3$		\\ \hline
C7:& \multicolumn{2}{c||}{$ |M_{jj}-M_{W^\pm}|< 30 \ \mbox{GeV}$}			& 17.5	& 726.2			& 54.5	& 26.2 & 810.9			\\ \hline
C8:& \multicolumn{2}{c||}{$ |M_{\ell \nu}-M_{W}|< 20 \ \mbox{GeV}$}		& 15.3	& 585.3			& 42.9	& 19.7 & 647.9			\\ \hline
\hline
\multicolumn{3}{||c||}{$S/B$ }   &	&	&	& &	2.4\%							\\\hline
\multicolumn{3}{||c||}{$S/\sqrt{B}$ with 100 fb$^{-1}$}   &  	&	& & 	& 6.1\\
\hline
\end{tabular}
\caption{ Cut flow of the cross sections for all  signals and backgrounds at the 14 TeV LHC. Conjugate processes are included here. \label{tab:14Tev_cutflow}}
\end{center}
\end{table}

\begin{itemize}
 \item {\bf $W^\pm h$ and $W^\pm A$ signal reconstruction}
 
\begin{enumerate}

\item {\bf Neutral Higgs candidate}: we search for a pair of $b$-tagged jets with minimum $\Delta R$. With this pair a neutral Higgs boson candidate is reconstructed.
In figure~\ref{fig:delr_bb} we display the $\Delta R$ separation between the two hardest $b$-jets, which are also those closer in phase space. Clearly, the signal distribution peaks 
at very low values of $\Delta R$ indicating a highly boosted Higgs boson decay which is the consequence of having a heavy charged Higgs state at source, as previously discussed. In order to make use of this feature, we thus reconstruct the Higgs bosons in our analysis from the pair of $b$-jets having minimum $\Delta R$. In figure~\ref{fig:INV_Mbb} we show the invariant mass distribution of a pair of $b$-jets with minimum $\Delta R$ separation. We see that, in the left panel of figure~\ref{fig:INV_Mbb}, as $M_h$ and $M_A$ are very close, the peaks corresponding to $h$ and $A$ get merged to give rise to a single fat $M_{bb}$ peak. In contrast, for $M_A=$ 130 GeV and 150 GeV, we can clearly see two different peaks corresponding to the two separate states  $h$ and $A$. Here, we further impose the following cuts on the invariant mass of the pair of $b$-jets: $|M_{b\bar b}-M_X|<15$ GeV, where $X=h,A$, to separate the $W^\pm h$ and $W^\pm A$ signals.

\item {\bf Top candidate}: since there are more than one $W^\pm$ boson, we choose one of them at a time and combine it with the remaining $b$-jet to find the closest invariant mass to the top-quark one with the requirement $$|M_{W^\pm b} - M_t|<30~ \mbox{GeV}.$$ The $W^\pm$ boson which provides the best solution is selected whereas the other is  retained to reconstruct the $H^\pm$ boson. In the left panel of the figure~\ref{fig:INV_MWh} we show the invariant mass distribution of the reconstructed $W^\pm$ plus $b$-jet and find that the distribution peaks at around the top mass, as expected, for both signal and backgrounds.


\item {\bf Charged Higgs candidate}: finally, we reconstruct the charged Higgs mass from the remaining $W^\pm$ and the reconstructed neutral Higgs $X$. In the right panel of  figure \ref{fig:INV_MWh} we show the invariant mass distribution of the $W^\pm$ and $X$ system. We find that the distribution peaks at the charged Higgs mass for the signal while
the backgrounds are efficiently  filtered out. 
\end{enumerate}

\begin{figure}[h!]
\begin{center}
\includegraphics[scale=0.3]{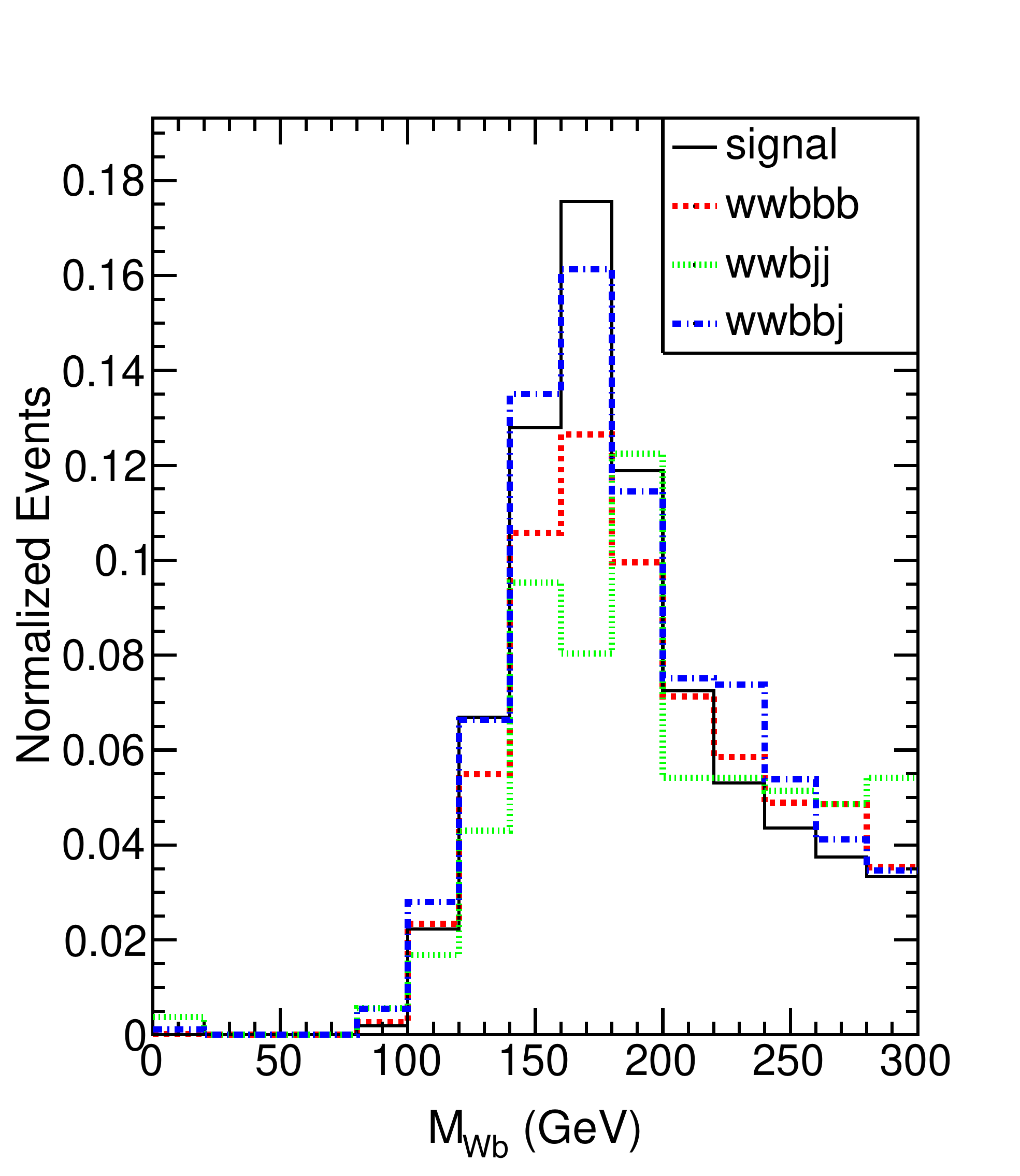}
\includegraphics[scale=0.3]{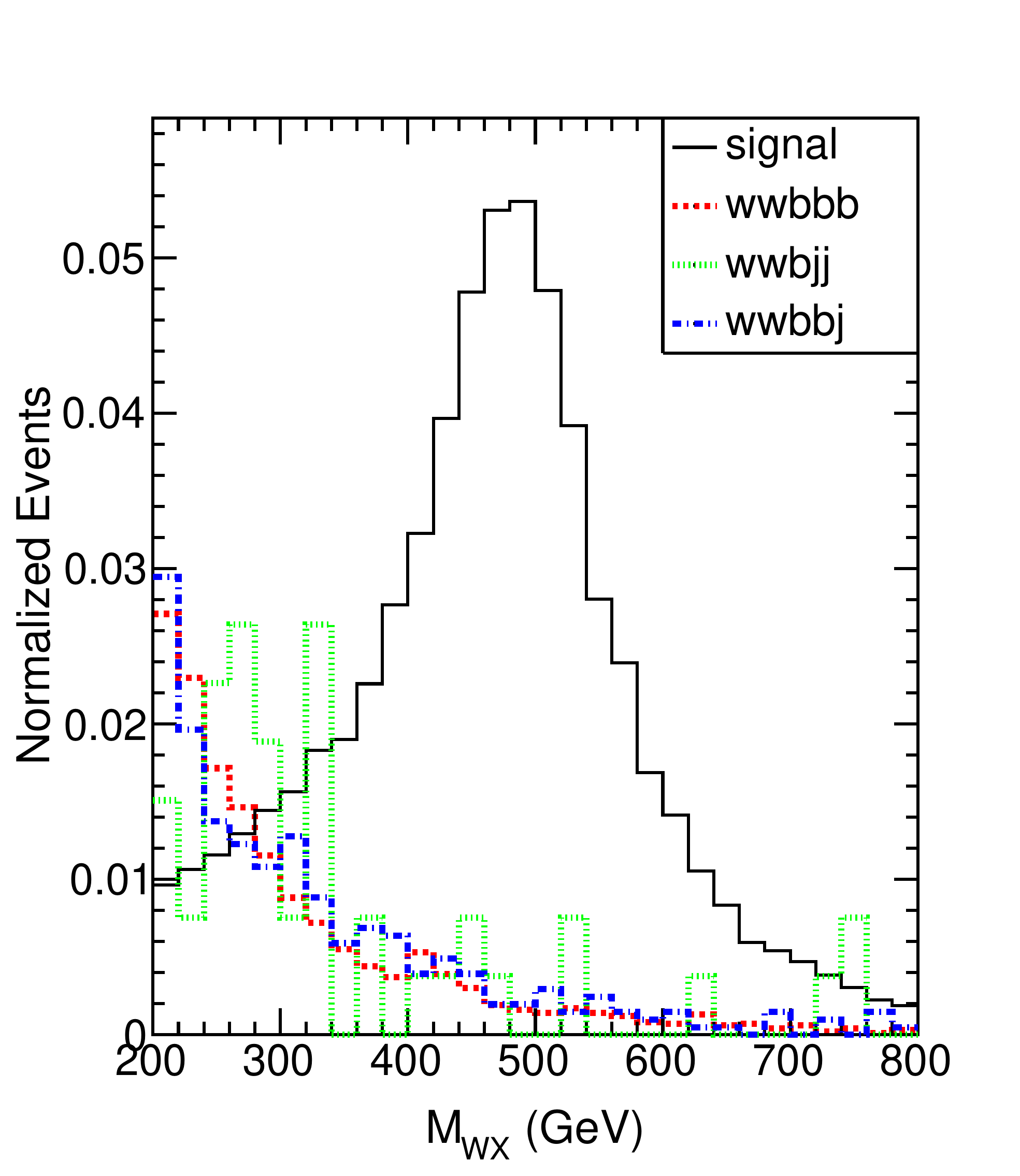}
\caption{\label{fig:INV_MWh} Invariant mass ($M_{Wb}$) of $W^\pm$ and remaining $b$-jet (left) and ($M_{WX}$) of the 
other $W^\pm$ and of the reconstructed $h,~A$ state (right) for the signal and backgrounds. }
\end{center}
\end{figure}

\begin{table}[h!]
\begin{center}
{\renewcommand{\arraystretch}{1.5}}
 \newcolumntype{C}[1]{>{\centering\let\newline\\\arraybackslash\hspace{0pt}}m{#1}} 
\begin{tabular}{ |C{0.85cm}C{2.5cm}C{1.75cm}| C{1.5cm} | C{1.5cm} | C{1.5cm}|C{1.5cm} |C{2.25cm} |}
\hline 
\multicolumn{3}{|c|}{\multirow{2}{*}{Cuts}}&\multicolumn{5}{c|}{$\sigma$ [fb] } \\\cline{4-8}
&&&Signal & $WWbbb$  & $WWbbj$  & $WWbjj$ & Total Background   \\\cline{1-8}
\hline
C9:& \multicolumn{2}{ c| }{ $ |M_{bb}-M_h|< 20 \ \mbox{GeV}$} 	 & 8.4  & 292.7 & 24.8 	& 11.4 & 328.9	 	\\ \hline
C10:& \multicolumn{2}{ c| }{ $ |M_{Wb}-M_{t}|< 30 \ \mbox{GeV}$} & 6.6	& 260.6 & 20.3	& 8.7  & 289.7		\\ \hline
C11:& \multicolumn{2}{c|}{ $ |M_{Wh}-M_{H^+}|< 100 \ \mbox{GeV}$}& 6.4	& 109.3	& 10.3	& 8.5  & 128.1		\\ \hline\hline
\multicolumn{3}{|c|}{$S/B$ }   &	&	&	&	&	5.4\%					\\\hline
\multicolumn{3}{|c|}{$S/\sqrt{B}$ with 100 fb$^{-1}$}   & 	&	& 	& & 5.9
 \\ \hline
\end{tabular}
\caption{ Cut flow of the cross sections for the $W^\pm X$ signals, $X={h,A}$, and backgrounds at the 14 TeV LHC. Conjugate processes are included here. \label{tab:14Tev_Wh}}
\end{center}
\end{table}

The final cut flow for this part is found in table \ref{tab:14Tev_Wh}. Rather standard luminosities are required for the extraction of these two (combined) signals.

\item {\bf $tb$ signal reconstruction}

\begin{enumerate}
 \item {\bf Top candidates}: from the two $W^\pm$s, we choose one and loop over the 3 $b$-jets. The $W b$ pair which has the invariant mass closest to the top-quark mass is selected. Then this process is repeated with the other $W^\mp$ and the remaining $b$-jets in order to find another top candidate. We select events satisfying $|M_{W^\pm b} - M_t|<30~ \mbox{GeV}$ for both top candidates. In figure~\ref{fig:INV_Mtb} (left), we show the invariant mass distribution $M_{Wb}$ of the reconstructed $W^\pm$ boson and $b$ jet. We show both the leptonic and hadronic top quarks reconstructed from the leptonic $W^\pm$ and hadronic $W^\pm$ bosons.
 
\item {\bf Charged Higgs  candidate}: with the two top quarks reconstructed, we select one and pair it with the remaining $b$-jets. The same process is repeated with the other top and the one with the invariant mass closest to charged Higgs mass, {\it viz.} $ |M_{tb}-M_{H^+}|< 100 ~\mbox{GeV}$ is kept in both cases. In figure~\ref{fig:INV_Mtb} (right), we show the invariant mass distribution of the reconstructed top quark and remaining $b$ jet.
\end{enumerate}

\begin{figure}[h!]
\begin{center}
\includegraphics[scale=0.3]{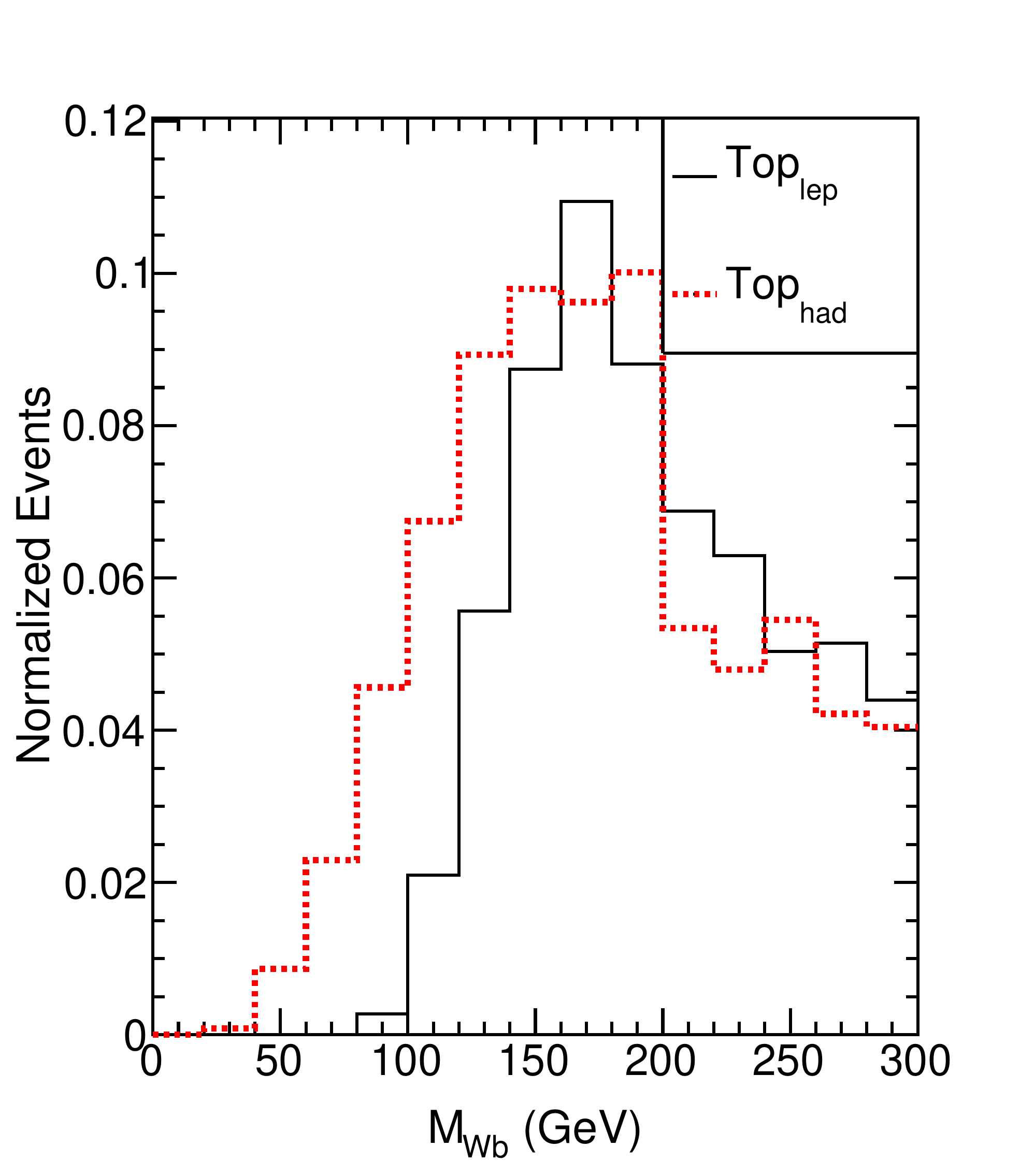}
\includegraphics[scale=0.3]{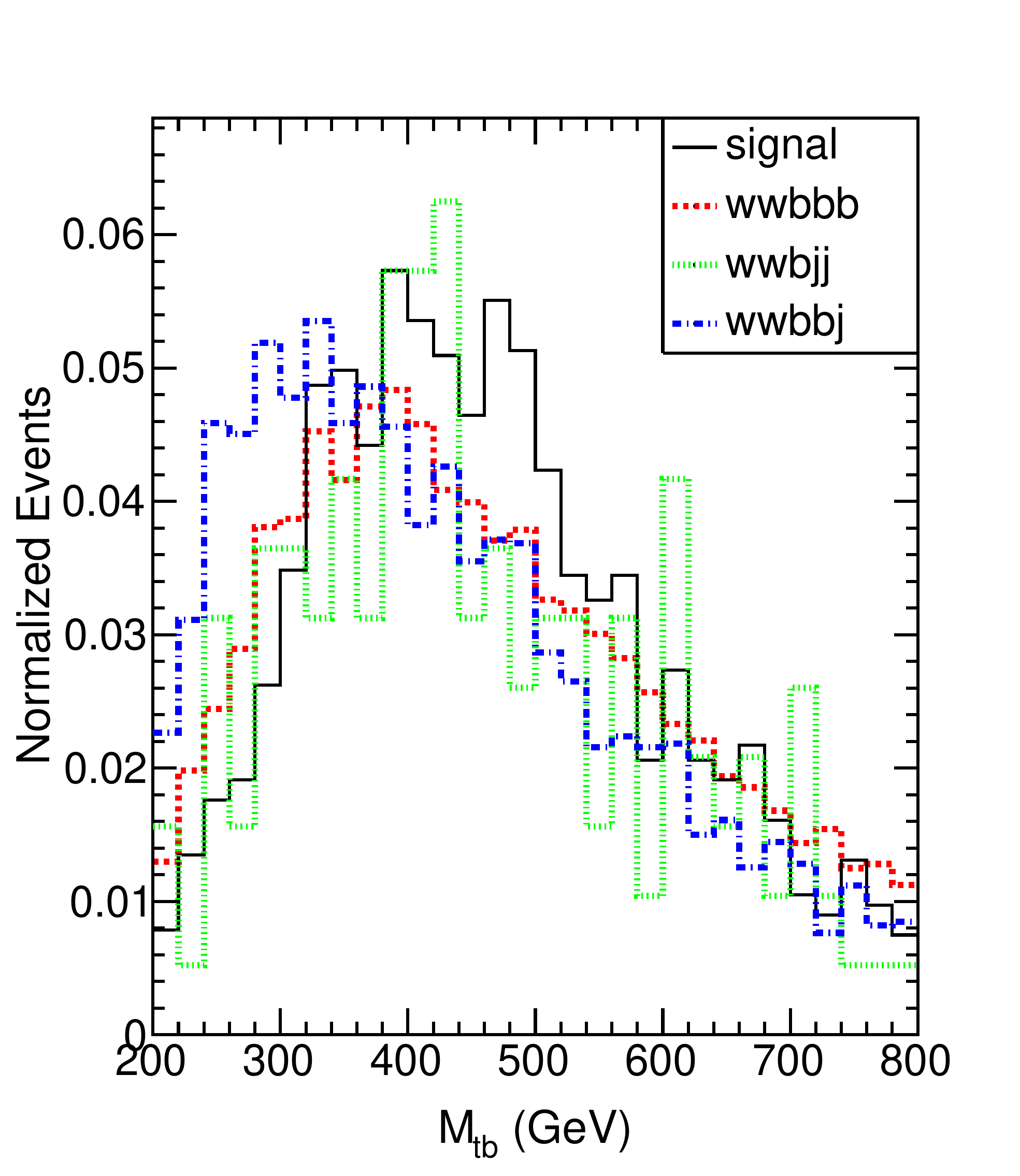}
\caption{\label{fig:INV_Mtb} Invariant mass of the reconstructed tops from the $W^\pm b$ distribution (left) and reconstructed charged Higgs boson mass from the $M_{tb}$ distribution (right) using one of the reconstructed top quarks and the remaining $b$ jet for the signal and backgrounds. }
\end{center}
\end{figure}

The final cut flow for this part is found in table \ref{tab:14Tev_tb}. Very large luminosities are required for the extraction of this signal.

\begin{table}[h!]
\begin{center}
{\renewcommand{\arraystretch}{1.5}}
 \newcolumntype{C}[1]{>{\centering\let\newline\\\arraybackslash\hspace{0pt}}m{#1}} 
\begin{tabular}{ |C{0.95cm}C{2.5cm}C{1.75cm}| C{1.5cm} | C{1.5cm} | C{1.5cm} |C{1.5cm} |C{2.25cm} |}
\hline 
\multicolumn{3}{|c|}{\multirow{2}{*}{Cuts}}&\multicolumn{5}{c|}{$\sigma$ [fb] } \\\cline{4-8}
&&&Signal & $WWbbb$  & $WWbbj$  & $WWbjj$ & Total Background    \\\cline{1-8}
\hline
C$9^\prime$:&  \multicolumn{2}{c|}{ $ |M_{Wb}-M_{t}|< 30 \ \mbox{GeV}$}   & 2.6	& 209.1 & 13.6	& 4.1 & 226.8	\\ \hline
C$10^\prime$:& \multicolumn{2}{c|}{ $ |M_{tb}-M_{H^+}|< 100 \ \mbox{GeV}$}& 1.4	& 175.3	& 8.9	& 3.2 & 187.4	\\ \hline\hline

\multicolumn{3}{|c|}{$S/B$ }   &	&	&	&	&		0.75\%			\\\hline
\multicolumn{3}{|c|}{$S/\sqrt{B}$ with 3000 fb$^{-1}$}   &  	&	& &	&5.7
 \\ \hline
\end{tabular}
\caption{ Cut flow of the cross sections for the $t\bar b$ signal and backgrounds at the 14 TeV LHC. Conjugate processes are included here. \label{tab:14Tev_tb}}
\end{center}
\end{table}

\end{itemize}

\end{itemize}

\section{Conclusions}

The 2HDM is the minimal extension of the SM that predicts the existence of charged Higgs bosons in the particle spectrum. A light charged Higgs state, below about 100 GeV, was already excluded by LEP~\cite{Abbiendi:2013hk}. After the LHC Run 1 the bounds have improved but they are $\tan \beta$ dependent and consequently we have now exclusion regions in the charged Higgs mass versus $\tan \beta$ plane. Although larger values of the masses are now excluded, there is a strong dependence on the Yukawa model type. A dedicated study for charged Higgs boson detection in all 2HDM types for a Higgs mass below the top-quark mass was performed in~\cite{Aoki:2011wd} for the 14 TeV LHC. The main conclusion was that, in the 2HDM-II, the whole parameter space would be probed for a light charged Higgs boson. For other types, probing the large $\tan \beta$ region has shown to be extremely hard if at all possible.

In this work we have instead focused on the heavy charged Higgs boson in the 2HDM-II. In fact, constraints from  $b \to s \gamma$ have raised the lower limit of a 2HDM-II charged Higgs boson mass to about 480 GeV. This raises the question of  whether such a heavy charged Higgs state can be detected during the current  LHC run. We have chosen a scenario where all possible decay channels are kept open. Since they all contribute to the most relevant signature, which is $WWbbb$, we have considered the simultaneous contribution of the different intermediate states  $W^\pm h$, $W^\pm A$, $W^\pm H$ (which is however subleading as we have taken $H$ to be SM-like) and $tb$. The main production cross section is associated production of a top-quark and a charged Higgs boson. This cross section is larger for either small or large $\tan \beta$. With all the theoretical and experimental constraints taken into account, only the low $\tan \beta$ region survives. Therefore we have chosen as our benchmark a value of $\tan \beta$ of order 1. The chances of finding a charged Higgs boson in fact degrade
considerably as $\tan \beta$ increases due to the dependence of the cross section upon it. We have finally shown that the prospects of detecting a heavy charged Higgs state with a mass of 500 GeV at the next LHC run are very good already for an integrated luminosity of 100 fb$^{-1}$ for the $W^\pm h$ and $W^\pm A$ modes while for the $tb$ one a tenfold increase in luminosity would be required \cite{Gianotti:2002xx}. Furthermore, we have also shown that it is possible to distinguish between the different intermediate states provided the scalar masses are sufficiently apart, although very high luminosities are required.  

Finally one should note that the benchmark proposed gives very similar results for the four Yukawa types of the 2HDM. For $\tan \beta$ of order 1 not only the cross sections are the same but also the main decay channels are similar because only $H^{\pm} \to tb$ depends on the Yukawa type and it has the same width in all four models for $\tan \beta$ small. 

In short, we believe to have set the stage for profitable analyses of the heavy mass region of the $H^\pm$ state in 2HDMs, that might eventually enable its discovery at the LHC, so we advocate ATLAS and CMS to follow the trail we have opened with more realistic experimental analyses. However, given the level of sophistication of our study,  we are confident that the former will corroborate the findings of the latter.

\section*{Acknowledgements}
SM is supported in part through the NExT Institute. SM and RS are supported by the grant H2020-MSCA-RISE-2014 no. 645722 (NonMinimalHiggs).

\end{document}